\documentclass[acmsmall,screen,authorversion]{acmart}
\usepackage{xcolor}
\usepackage[export]{adjustbox}
\usepackage{tabularx}
\usepackage{algorithm}
\usepackage{algpseudocode}
\usepackage{multirow}

\AtBeginDocument{%
  \providecommand\BibTeX{{%
    \normalfont B\kern-0.5em{\scshape i\kern-0.25em b}\kern-0.8em\TeX}}}

\setcopyright{acmlicensed}
\copyrightyear{2022}
\acmYear{2022}
\acmDOI{10.1145/3568392}

\acmJournal{FACMP}
\acmVolume{0}
\acmNumber{0}
\acmArticle{0}
\acmMonth{0}

\begin{document}

\title{Auditing YouTube's Recommendation Algorithm for Misinformation Filter Bubbles}


\author{Ivan Srba}
\affiliation{%
  \institution{Kempelen Institute of Intelligent Technologies}
  %\streetaddress{Mlynske nivy 5}
  \city{Bratislava}
  \country{Slovakia}
}
\email{ivan.srba@kinit.sk}
\orcid{0000-0003-3511-5337}

\author{Robert Moro}
\affiliation{%
  \institution{Kempelen Institute of Intelligent Technologies}
  %\streetaddress{Mlynske nivy 5}
  \city{Bratislava}
  \country{Slovakia}
}
\email{robert.moro@kinit.sk}
\orcid{0000-0002-3052-8290}

\author{Matus Tomlein}
\affiliation{%
  \institution{Kempelen Institute of Intelligent Technologies}
  %\streetaddress{Mlynske nivy 5}
  \city{Bratislava}
  \country{Slovakia}
}
\email{matus.tomlein@kinit.sk}
\orcid{0000-0002-9960-700X}

\author{Branislav Pecher}
\affiliation{%
  \institution{Faculty of Information Technology, Brno University of Technology}
  %\streetaddress{Mlynske nivy 5}
  \city{Brno}
  \country{Czechia}
}
\additionalaffiliation{%
  \institution{Kempelen Institute of Intelligent Technologies}
  %\streetaddress{Mlynske nivy 5}
  \city{Bratislava}
  \country{Slovakia}
}
\email{branislav.pecher@kinit.sk}
\orcid{0000-0003-0344-8620}

\author{Jakub Simko}
\affiliation{%
  \institution{Kempelen Institute of Intelligent Technologies}
  %\streetaddress{Mlynske nivy 5}
  \city{Bratislava}
  \country{Slovakia}
}
\email{jakub.simko@kinit.sk}
\orcid{0000-0003-0239-4237}

\author{Elena Stefancova}
\affiliation{%
  \institution{Kempelen Institute of Intelligent Technologies}
  %\streetaddress{Mlynske nivy 5}
  \city{Bratislava}
  \country{Slovakia}
}
\email{elena.stefancova@kinit.sk}
\orcid{0000-0001-8683-939X}

\author{Michal Kompan}
\affiliation{%
  \institution{Kempelen Institute of Intelligent Technologies}
  %\streetaddress{Mlynske nivy 5}
  \city{Bratislava}
  \country{Slovakia}
}
\additionalaffiliation{%
  \institution{slovak.AI}
  %\streetaddress{Ilkovicova 2}
  \city{Bratislava}
  \country{Slovakia}
}
\email{michal.kompan@kinit.sk}
\orcid{0000-0002-4649-5120}

\author{Andrea Hrckova}
\affiliation{%
  \institution{Kempelen Institute of Intelligent Technologies}
  %\streetaddress{Mlynske nivy 5}
  \city{Bratislava}
  \country{Slovakia}
}
\email{andrea.hrckova@kinit.sk}
\orcid{0000-0001-9312-6451}

\author{Juraj Podrouzek}
\affiliation{%
  \institution{Kempelen Institute of Intelligent Technologies}
  %\streetaddress{Mlynske nivy 5}
  \city{Bratislava}
  \country{Slovakia}
}
\email{juraj.podrouzek@kinit.sk}
\orcid{0000-0002-9691-0310}

\author{Adrian Gavornik}
\affiliation{%
  \institution{Kempelen Institute of Intelligent Technologies}
  %\streetaddress{Mlynske nivy 5}
  \city{Bratislava}
  \country{Slovakia}
}
\email{adrian.gavornik@intern.kinit.sk}
\orcid{0000-0001-5678-0083}

\author{Maria Bielikova}
\affiliation{%
  \institution{Kempelen Institute of Intelligent Technologies}
  %\streetaddress{Mlynske nivy 5}
  \city{Bratislava}
  \country{Slovakia}
}
\additionalaffiliation{%
  \institution{slovak.AI}
  %\streetaddress{Ilkovicova 2}
  \city{Bratislava}
  \country{Slovakia}
}
\email{maria.bielikova@kinit.sk}
\orcid{0000-0003-4105-3494}

\renewcommand{\shortauthors}{Srba, et al.}

\definecolor{DarkGreen}{RGB}{36,135,33}
\begin{abstract}
  In this paper, we present results of an auditing study performed over YouTube aimed at investigating how fast a user can get into a misinformation filter bubble, but also what it takes to ``burst the bubble'', i.e., revert the bubble enclosure. We employ a sock puppet audit methodology, in which pre-programmed agents (acting as YouTube users) delve into misinformation filter bubbles by watching misinformation promoting content. Then they try to burst the bubbles and reach more balanced recommendations by watching misinformation debunking content. We record search results, home page results, and recommendations for the watched videos. Overall, we recorded 17,405 unique videos, out of which we manually annotated 2,914 for the presence of misinformation. The labeled data was used to train a machine learning model classifying videos into three classes (promoting, debunking, neutral) with the accuracy of 0.82. We use the trained model to classify the remaining videos that would not be feasible to annotate manually.

Using both the manually and automatically annotated data, we observe the misinformation bubble dynamics for a range of audited topics. Our key finding is that even though filter bubbles do not appear in some situations, when they do, it is possible to burst them by watching misinformation debunking content (albeit it manifests differently from topic to topic). We also observe a sudden decrease of misinformation filter bubble effect when misinformation debunking videos are watched after misinformation promoting videos, suggesting a strong contextuality of recommendations. Finally, when comparing our results with a previous similar study, we do not observe significant improvements in the overall quantity of recommended misinformation content.
\end{abstract}




\maketitle

\section{Introduction}
\label{sec:intro}

In this paper, we investigate the \emph{misinformation filter bubble} creation and bursting on YouTube\footnote{This paper is an extended version of a paper entitled ``An Audit of Misinformation Filter Bubbles on YouTube: Bubble Bursting and Recent Behavior Changes''~\cite{tomlein_audit_2021}, which has been awarded the Best Paper Award at the Fifteenth ACM Conference on Recommender Systems (RecSys ’21).}. The role of very large online platforms (especially social networking sites, such as Facebook, Twitter, or YouTube) in dissemination and amplification of misinformation has been widely discussed and recognized in recent years by researchers, journalists, policymakers, and representatives of the platforms alike~\cite{Ribeiro2020,Spinelli2020,gpai_responsible_ai_2021,eu_tackling_disinfo_2018,eu_multi-dimensional_2018,coalition_to_fight_digital_deception_trained_2021,hagey_horwitz_facebook_2021}. The platforms are blamed for promoting sensational, attention-grabbing, or polarizing content through the use of personalized recommendation algorithms (resulting from their mode of operation based on monetizing users' attention~\cite{zuboff2019age, vaidhyanathan2018antisocial}). To tackle this issue, the platforms have (on the European level) committed to implement a range of measures stipulated in the Code of Practice on Disinformation~\cite{code_of_practice_2018}. However, the monitoring of the platforms' compliance and the progress made in this regard has proved difficult~\cite{erga_report_2020}. One of the problems is a lack of effective public oversight in the form of \emph{internal audits} of the platforms' personalized algorithms that could directly quantify the impact of disinformation as well as the measures taken by the platforms.

This lack has been partially compensated by \emph{external black-box auditing studies} performed by the researchers, such as~\cite{Hussein2020, Papadamou2020, Spinelli2020, AbulFottouh2020, Ribeiro2020, ballardConspiracyBrokersUnderstanding2022} that aimed to quantify the portion of misinformative content being recommended on social media platforms. With respect to YouTube, which is the subject of the audit presented in this paper, previous works investigated how a user can enter a filter bubble. Multiple studies demonstrated that watching a series of misinformative videos strengthens the further presence of such content in recommendations~\cite{Hussein2020, Papadamou2020, AbulFottouh2020}, or that following a path of the ``up next'' videos can bring the user to a very dubious content~\cite{Spinelli2020}. However, no studies have covered \emph{if}, \emph{how} or with what \emph{``effort''} can the user ``burst'' (lessen) the bubble. More specifically, they have not investigated what type of user's watching behavior (e.g., switching to credible news videos or conspiracy debunking videos) would be needed to lessen the amount of misinformative content recommended to the user. Such knowledge would be valuable not just for the sake of having a better understanding of the inner workings of YouTube's personalization, but also to improve the social, educational, or psychological strategies for building up resilience against misinformation.

Our work extends the prior works by researching this important aspect. To do so, we employ a \emph{sock puppet} auditing methodology~\cite{bandy_problematic_2021,Sandvig2014Audits}. We simulate user behavior on the YouTube platform, record platform responses (search results, home page results, recommendations) and manually annotate their sample for the presence of misinformative content. Using the manual annotations, we train a machine learning model to predict labels for the remaining recommended videos that would be impractical to annotate manually due to their large volume. Then, we quantify the \emph{dynamics} of misinformation filter bubble creation and also of bubble bursting, which is the novel aspect of the study.

The main contributions of this work are threefold. \textbf{As the first contribution}, this paper reports on the behavior of YouTube's personalization in a situation when a user with misinformation promoting watch history (i.e., with a developed misinformation filter bubble) starts to watch content debunking the misinformation (in an attempt to burst that misinformation filter bubble). The key finding is that watching misinformation debunking videos (such as credible news or scientific content) generally improves the situation (in terms of recommended items or search result personalization), albeit with varying effects and forms, mainly depending on the particular misinformation topic. 

Complementing manual labels with automatically predicted ones (using our trained machine learning model) allowed us to inspect not only difference at the specific points in time (the state at the beginning of the study vs. the state after obtaining a watch history of misinformation promoting videos vs. the state after watching the misinformation debunking content), but also the \emph{dynamics} of misinformation filter bubble creation and bursting throughout the whole duration of the study. Thus, \textbf{as the second contribution}, we provide a so-far unexplored deeper insight into misinformation filter bubble dynamics, since a continuous evaluation of the proportions of misinformation promoting and debunking videos has not been covered, to the best of our knowledge, by any of the existing auditing studies yet. The key finding is that there is a sudden increase in the number of debunking videos after the first watched debunking video, suggesting a strong contextuality of the YouTube's personalization algorithms. We observe this consistently for both the home page results and the recommendations for most examined misinformation topics.

Lastly, part of this work is a replication of the prior works, most notably the work of Hussein et al.~\cite{Hussein2020} who also investigated the creation of misinformation filter bubbles using user simulation. We aligned our methodology with Hussein's study: we re-used Hussein's seed data (topics, queries, and also videos, except those which have been removed in the meantime), used similar scenarios and the same data annotation scheme. As a result, we were able to directly compare the outcomes of both studies, Hussein's and ours, on the number of observed misinformative videos present in recommendations or search results. \textbf{As the third contribution}, we report changes in misinformation video occurrences on YouTube, which took place since the study of Hussein et al.~\cite{Hussein2020} (mid-2019). Due to the ongoing YouTube's efforts to improve their recommender systems and policies (e.g., by removing misinformative content or preferring credible sources)~\cite{YouTube2020policies,YouTube2021perspective}, we expected to see less filter bubble creation behavior than Hussein et al. While we in general observe low overall prevalence of misinformation in several topics, there is still a room for improvement. More specifically, we observe worse situation regarding the topics of vaccination and (partially) 9/11 conspiracies and some improvements (less misinformation) for moon landing or chemtrails conspiracies. In addition, we replicated, to a lesser extent, the works of Hou et al.~\cite{Hou2019} and Papadamou et al.~\cite{Papadamou2020}. We reused their classification models, which we adapted and applied on our collected data. Finally, we used the better performing Papadamou's model~\cite{Papadamou2020} to predict the labels for the videos that were not labeled manually.

To ensure future replicability of our study and reproducibility of the presented results, the implementation of the experimental infrastructure, collected data, manual annotations as well as the analytical notebooks are all publicly available on GitHub\footnote{\url{https://github.com/kinit-sk/yaudit-recsys-2021}}. However, to mitigate the identified ethical risks associated with a possibility of mislabeling a video as promoting misinformation (cf. Section~\ref{sec:methodology:ethics}), we do not publish the labels predicted by our trained machine learning model, only their aggregated numbers necessary to reproduce the results. Nevertheless, since we train and use models published in prior works~\cite{Hou2019,Papadamou2020}, it is also possible to replicate this part of our study. In addition, we do not publish copyrighted data (video title, description, etc.). However, this metadata can be downloaded through YouTube API.  

The rest of the paper is structured as follows. Section~\ref{sec:background} looks at the definitions of a filter bubble, echo chambers, and misinformation and uses them to define a misinformation filter bubble which is the focus of this work. In Section~\ref{sec:relwork}, we analyze different types of auditing studies and what the previous works examined with respect to social media (YouTube in particular) and misinformation. In Section~\ref{sec:methodology}, we describe in detail our research questions and study methodology, including manual as well as automatic annotation of collected data. We present the results of a training of a machine learning model for automatic annotation of videos and compare these results with the relevant related works. We also discuss identified ethical issues and how we addressed them in our research. The results of the audit itself are presented in Section~\ref{sec:results}. Finally, the implications of the results and the conclusions are discussed in Section~\ref{sec:conclusions}. 
\section{Background: Filter Bubbles and Misinformation}
\label{sec:background}

% Echo chambers

In online social media, a phenomenon denoted as virtual \emph{echo chambers} refers to situations in which the same ideas are repeated, mutually confirmed and amplified in relatively closed homogeneous groups. Echo chambers may finally contribute into increased polarization and fragmentation of the society~\cite{journals/corr/ZolloNVBMSCQ15,sunstein1999law}. Tendency of users to join and remain in echo chambers can be intentional (also denoted as self-selected personalization \cite{borgesiusShouldWeWorry2016a}) and explained by their selective exposure (focusing on information that is in accordance with one’s worldview) or confirmation bias (reinforcing one's pre-existing beliefs)~\cite{bessi2016personality}. To some extent such users' behavior is a natural human defense against information overload~\cite{mutz2011communication}. However, enclosure of users into echo chambers can be also caused and amplified by adaptive systems (even without users' intention) resulting into the so-called \emph{filter bubble} effect (also denoted as pre-selected personalization~\cite{borgesiusShouldWeWorry2016a}). This effect has serious ethical implications---users are often unaware of the existence of filter bubbles, as well as of the information that was filtered out.

% Filter bubbles

Filter bubbles were firstly recognized by Pariser~\cite{pariser2011filter} in 2011 as a state of intellectual isolation caused by algorithms that personalize users' online experiences hence exposing users to information and opinions that conform to and reinforce their own beliefs. Filter bubbles quickly became the target of theoretical as well as empirical research efforts from multiple perspectives, such as: 1) exploring characteristics of filter bubbles and identification of circumstances of their creation~\cite{liuInteractionPoliticalTypology2021}; 2) modelling/quantifying the filter bubble effect~\cite{aridorDeconstructingFilterBubble2020,jiangDegenerateFeedbackLoops2019}; and 3) discovering strategies how to prevent or ``burst'' filter bubbles~\cite{chrysanthouBurstingBubbleTool2020}. 

The ambiguity and difficult operationalization of the original Pariser's definition of filter bubbles led to its different interpretations; inconsistent or even contrasting findings; and finally, also to low generalizability across the studies~\cite{michielsWhatAreFilter2022}. For these reasons, the operationalized, systematically and empirically verifiable definition of the filter bubble has been recently proposed in~\cite{michielsWhatAreFilter2022} as follows: ``A technological filter bubble is a decrease in the diversity of a user’s recommendations over time, in any dimension of diversity, resulting from the choices made by different recommendation stakeholders.'' Based on this definition, authors also stress the criteria on studies addressing the filter bubble effect---they must consider the diversity of recommendations and measure a decrease in diversity over time. In this work, we proceed from this definition and also meet the stated criteria.

% Misinformation

We are interested specifically in filter bubbles that are determined by the presence of content spreading disinformation or misinformation. \emph{Disinformation} is a ``false, inaccurate, or misleading information designed, presented and promoted to intentionally cause public harm or for profit''~\cite{eu_multi-dimensional_2018}, while \emph{misinformation} is a false or inaccurate information that is spread regardless of an intention to deceive. In this work, we use a broader term of misinformation since we are interested in any kind of false or inaccurate information regardless of intention behind its creation and spreading (in contrast to the most current reports and EU legislation that opt for the term disinformation and thus emphasize the necessary presence of intention). Due to significant negative consequences of misinformation on our society (especially during the ongoing COVID-19 pandemic), tackling misinformation attracted also a plethora of research works (see~\cite{Zannettou2019,Zhou2020,chorasAdvancedMachineLearning2021} for recent surveys). The majority of such research focuses on various characterization studies~\cite{simko2021study} or detection methods~\cite{pecher2021fireant,srba2019monant}.

% Misinformation filter bubbles

We denote filter bubbles that are characterized by the increased prevalence of such misinformative content as \emph{misinformation filter bubbles}. They are states of intellectual isolation in false beliefs or manipulated perceptions of reality. Following the adopted definition, filter bubbles in general are characterized by the decrease in any dimension of diversity. We can broadly distinguish three types of diversity~\cite{michielsWhatAreFilter2022}: structural, topical and viewpoint diversity.  Misinformation filter bubbles can be considered as a special case of a decrease of viewpoint diversity in which the viewpoints represented are provably false. Analogically to \emph{topical} filter bubbles, misinformation filter bubbles can be characterized by a high homogeneity of recommendations/search results that share the same positive stance towards misinformation. In other words, the content adaptively presented to a user in a misinformation filter bubble supports one or several false claims/narratives. While topical filter bubbles are not necessary undesirable (they may be intended and even positively perceived by users~\cite{burbachBubbleTroubleStrategies2019,gpai_responsible_ai_2021}), misinformation filter bubbles are by definition more problematic and cause an indisputable negative effects~\cite{fernandezAnalysingEffectRecommendation2021,coalitiontofightdigitaldeceptionTrainedDeceptionHow2021,gpai_responsible_ai_2021}.

Reflecting the adopted definition of the filter bubble~\cite{michielsWhatAreFilter2022}, we do not consider misinformation filter bubble as a binary state at a single moment (i.e., following the current recommended items, a user is/is not in the misinformation filter bubble), but as the interval measure reflecting how deep inside the bubble the user is. Such a measure is determined by the proportion of misinformative content and calculated at different points in time. This definition and operationalization of filter bubbles emphasizes our second contribution in so-far unexplored deeper insight into the dynamics of filter bubbles since we calculated the proportion of misinformative content not only in the selected time points but continuously over the whole duration of the study.

% Current social media efforts 

To prevent misinformation and misinformation filter bubbles, social media conduct various countermeasures. These are usually reactions to public outcry or are required by legislation, e.g., EU's Code of Practice on Disinformation~\cite{code_of_practice_2018}. Currently, the effectiveness of such countermeasures is evaluated mainly by self-evaluation reports. This approach has been, however, already recognized as insufficient due to a lack of evidence and objectivity since social media are reluctant to provide access to their data for independent research~\cite{erga_report_2020}. In addition, the commercial aims of social media may be contradicting pro-social interests, as also revealed by the recent whistleblowing case Facebook Files\footnote{Facebook Files denote a leak of internal documents revealing that the company was aware of negative societal impact caused by the platform, including its algorithms, like spreading and preferring harmful or controversial content. At the same time, documents showed that the company's reactions on such caused real-world harms were not sufficient.}~\cite{horwitzFacebookFilesWall2021}. The verification of countermeasures is further complicated by interference of psychological factors. For example, some researchers argue that users' intentional self-selected personalization is more influential than algorithms' pre-selected personalization when it comes to intellectual isolation~\cite{del2016spreading, bakshy2015exposure}.

% Audits

An alternative solution towards responsible and governed AI in social media and eliminating its negative social impact is employment of independent \emph{audits}. Such audits, which are carried out by an external auditor independent from the company developing the audited AI algorithm, are envisaged also in the proposal of an upcoming EU legislation~\cite{europeancommissionProposalRegulationEuropean2020}. Nevertheless, the auditing studies on the intersection of filter bubbles and misinformation (i.e., studies on misinformation filter bubbles), such as the one presented in this paper, are still relatively rare.
\section{Related work: Audits of Adaptive Systems}
\label{sec:relwork}

In this context, an audit is a systematic statistical probing of an online platform, used to uncover socially problematic behavior underlying its algorithms~\cite{Sandvig2014Audits,Hussein2020}. 
Algorithmic audits can be conducted in two different settings. Internal (white-box) audits may utilize a direct access to the recommender system (algorithm and data) and thus require a cooperation between an independent auditor and the platform operating the recommender system. External (black-box) audits are performed without the detailed information and access to the internal workings of the recommender system, and thus are limited to publicly available data obtained via user interface or API. External audits may suffer from methodological problems, especially limited possibilities to evaluate causal hypotheses about the effects of different variants of recommender systems on users~\cite{gpai_responsible_ai_2021}. Internal audits may naturally lead to more precise results, nevertheless, their execution is not currently possible for independent researchers (we do not expect that the situation will change significantly in future even when the new EU legislation~\cite{europeancommissionProposalRegulationEuropean2020} will be adopted~\cite{mesarcik_analysis_2022}). Fernández et al. \cite{fernandezAnalysingEffectRecommendation2021} attempted to overcome this limitation and performed simulations of the effect of some of the most popular recommendation algorithms on the spread of misinformation utilizing Twitter data. Such recommender systems are, however, very probably too distinct from the algorithms utilized by social media platforms. Therefore, the applicability of findings on real-world social media recommender systems is questionable. Auditing filter bubbles caused by the social media recommender systems are thus performed almost solely externally. 

From another perspective, algorithmic auditing can address four categories of problematic machine behavior~\cite{bandy_problematic_2021}: discrimination, distortion, exploitation, and misjudgement. Auditing filter bubbles caused by social media recommender systems falls into the distortion category (i.e., a behaviour that distorts or obscures an underlying reality).

External audits examining distortion category of problematic behavior come in multiple forms~\cite{bandy_problematic_2021} as defined by the taxonomy introduced by Sandvig et al. ~\cite{Sandvig2014Audits}. Commonly used \emph{scraping audits}, which collect data by generating requests to API or web interface, allow only very limited simulation of user behavior. \emph{Crowdsourcing audits} and \emph{sockpuppeting audits}, which can replicate user behavior more precisely and eliminate confound factors better, are more suitable to investigate the effect of (misinformation) filter bubbles; nevertheless, they have been researched to a lesser extent.

% Crowdsourcing audits

Crowdsourcing audit studies are conducted using real user data. Hannak et al.~\cite{Hannak2013} recruited Mechanical Turk users to run search queries and collected their personalized results. Silva et al.~\cite{Silva2020} developed a browser extension to collect personalized ads with real users on Facebook. Shen et al.~\cite{shenEverydayAlgorithmAuditing2021} introduced an idea of everyday algorithm auditing, in which users detect, understand, and interrogate problematic machine behaviors via their day-to-day interactions with algorithmic systems. However, such auditing methodology suffers from a lack of isolation (users may be influenced by additional factors, such as confirmation bias). Moreover, uncontrolled environment makes comparisons difficult or unfeasible; it is difficult to keep users active; such audits also raise several privacy issues.

% Sockpuppeting audits

Sockpuppeting audits solve these problems by employing non-human bots that impersonate the behavior of users in a predefined controlled way~\cite{Sandvig2014Audits}. To achieve representative and meaningful results in sockpuppeting audits, researchers need to tackle several methodological challenges~\cite{Hussein2020}. First is the selection of appropriate seed data, which are needed to setup user profiles and prescribe user actions executed during the audit (e.g., the initial activity of bots, search queries). Second, the experimental setup must measure the real influence of the investigated phenomena. At the same time, it must minimize confounding factors and noise (such as of name, gender or geolocation~\cite{Hannak2013}). Another challenge is how to appropriately label the presence of the audited phenomena (expert-based/crowdsourced~\cite{Hussein2020,Silva2020} or automatic labeling~\cite{Papadamou2020} can be employed).  

% Additional distinction features of audits

Audits can be further distinguished by the online platform they are applied on (social networking sites~\cite{Silva2020,Papadamou2020,Hussein2020,haroonYouTubeGreatRadicalizer2022,ballardConspiracyBrokersUnderstanding2022}, search engines~\cite{Metaxa2019,Le2019,Robertson2018}, e-commerce sites~\cite{Juneja2021}), by adaptive systems being investigated (recommendations~\cite{Hussein2020,Spinelli2020,Papadamou2020}, up-next recommendation~\cite{Hussein2020}, search results~\cite{Papadamou2020,Hussein2020,Le2019,Metaxa2019,Robertson2018}, autocomplete~\cite{Robertson2018}, advertisement system~\cite{ballardConspiracyBrokersUnderstanding2022}) and by phenomena being studied (misinformation~\cite{Hussein2020,Papadamou2020}, political/ideological bias~\cite{Le2019,Metaxa2019,haroonYouTubeGreatRadicalizer2022}, political ads~\cite{Silva2020}). In our study, we focus specifically on misinformation filter bubbles in the context of the online video platform YouTube and its recommender and search system. As argued by Spinelli et al.~\cite{Spinelli2020}, YouTube is an important case to study as a significant source of socially-generated content and because of its opaque recommendation policies.
% YouTube and YouTube RS
Some information about the inner workings of YouTube's adaptive systems are provided by research papers published at RecSys conference~\cite{Covington2016,Zhao2019} or blogs~\cite{YouTube2020policies} published directly by the platform, nevertheless, a detailed information is unknown. Therefore, we feel a need to conduct independent auditing studies on undesired phenomena like unintended creation of misinformation filter bubbles.

% YouTube related audits

The existing studies confirmed the effects of filter bubbles in YouTube recommendations and search results. Spinelli et al.~\cite{Spinelli2020} found that chains of recommendations lead away from reliable sources and toward extreme and unscientific viewpoints. Similarly, Ribeiro et al.~\cite{Ribeiro2020} concluded that YouTube's recommendation contributes to further radicalization of users and found paths from large media channels to extreme content through recommendation. Abul-Fottouh et al.~\cite{AbulFottouh2020} confirmed a homophily effect in which anti-vaccine videos were more likely to recommend other anti-vaccine videos than pro-vaccine ones and vice versa. In a recent work, Haroon et al.~\cite{haroonYouTubeGreatRadicalizer2022} utilized sock puppets to determine the presence of ideological bias (i.e., the alignment of recommendations with users’ ideology), its magnitude (i.e., whether the users are recommended an increasing number of videos aligned with their ideology), and radicalization (i.e., whether the recommendations are progressively more extreme). Furthermore, a bottom-up intervention to minimize ideological bias in recommendations was designed and evaluated. Ballard et al.~\cite{ballardConspiracyBrokersUnderstanding2022} analyzed how ads shown on conspiracy content differ from ads on mainstream videos and thus evaluated YouTube's algorithms for advertisement selection and demonetization. Obtained results revealed that conspiracy videos have fewer, but at the same time lower-quality and more predatory ads than mainstream videos. Authors also conclude that the difference in advertising quality suggests that YouTube’s advertising platform may be assisting predatory advertisers to identify potential victims.

% YouTube misinformation filter bubble audits

Recently, we can observe first audits focused specifically on misinformation filter bubbles in the case of YouTube recommender systems. Hussein et al.~\cite{Hussein2020} and Papadomou et al.~\cite{Papadamou2020} found that YouTube mitigates pseudoscientific content in some handpicked topics such as COVID-19. Hussein et al.~\cite{Hussein2020} found that demographics and geolocation (within the US) affect personalization only after having acquired some watch history. These studies provide evidence of the existence and properties of misinformation filter bubbles on YouTube.

% Automated annotation methods

Above-mentioned auditing studies took two different approaches to determine the presence of investigated phenomenon in the recommended content. At first, some studies rely on manual (expert) annotation~\cite{Hussein2020}. This approach can potentially lead to more precise labels. Annotation process may be, however, influenced by subjectivity of individual human experts and, therefore, each recommended item should be assigned a final label only in case of agreement of multiple experts. In the end, such approach is very time-consuming and, therefore, the amount of labeled data is rather small. Secondly, some studies use automatic annotations by means of various heuristics (e.g., analysis of user feedback~\cite{haroonYouTubeGreatRadicalizer2022}) or a machine learning model (e.g., a classification model~\cite{Papadamou2020}). Applying machine learning models allows to annotate a significantly larger amounts of data. Nevertheless, it still requires a labeled subset of data (labeled either by experts or by crowdsourcing) to create a necessary training set, and at the same time, machine learning predictions may introduce some level of noise due to their natural imperfection.

From the properties that remain uninvestigated, we specifically address two. First, the adaptive systems used by YouTube are in continuous development and improvement. Information on how YouTube proceeds in countering misinformation is needed. Second, while the existing studies focused on misinformation filter bubble creation, there is not the same level of insight on the inverse process---filter bubble bursting. The online survey performed in~\cite{burbachBubbleTroubleStrategies2019} revealed that the majority of respondents (63\%) are aware that they are affected by the filter bubble, while fewer participants (31\%) also deliberately take action against the filter bubble effect. Investigation of bubble bursting strategies can help not only such users who intentionally want to get more diverse/balanced recommendation mix, but also to get a better understanding how recommender system internally works (what can be valuable also for social media platforms themselves). Such insight can finally help to improve the design of recommender systems (e.g., how exactly to prioritize credible sources).
\section{Study design and methodology}
\label{sec:methodology}

To investigate the dynamics of bursting out of a misinformation filter bubble, we conducted an external agent-based  sockpuppeting audit study. The study took place on YouTube, but its methodology and implementation can be generalized to any adaptive service, where recommendations can be user-observed.

In the study, we let a series of agents (bots) pose as YouTube users. The agents performed pre-defined sequences of video watches and query searches. They also recorded items they saw: recommended videos, search results, and videos shown at the home page. The pre-defined actions were designed to first \emph{invoke the misinformation filter bubble effect} by purposefully watching videos with (or leaning towards) misinformative content. Then, agents tried to \emph{mitigate the bubble effect} by watching videos with trustworthy (misinformation debunking) content. Between their actions, the agents were idle for some time to prevent possible carry-over effects. The degree of how deep inside a bubble the agent is was observed through the number and rank of misinformative videos offered to them.

The secondary outcome is the partial replication of a previous study done by Hussein et al.~\cite{Hussein2020} (denoted hereafter as the \emph{reference study}). This replication allowed us to draw direct comparisons between quantities of misinformative content that agents encountered during our study (conducted in March 2021) and during the reference study (conducted in mid-2019).

\subsection{Research Questions and Hypotheses}
\label{sec:methodology:rqs}

\textbf{RQ1 (comparison to the reference study):} \emph{Has YouTube's personalization behavior changed with regards to misinformative videos since the reference study?} In particular, we seek to validate the following hypothesis:
\begin{itemize}
    \item \textbf{H1.1:} We will observe a weaker misinformation filter bubble effect, when comparing the state after constructing a promoting watch history with the results of the reference study in both search and recommendations.
\end{itemize}

\textbf{RQ2 (bubble bursting dynamics):} \emph{How does misinformation filter bubble effect change, when debunking videos are watched?} The ``means of bubble bursting'' would be implicit user feedback---watching misinformation debunking videos. In particular, we seek to validate the following hypotheses:

\begin{itemize}
    \item \textbf{H2.0:} Watching videos that promote misinformation leads to their increased presence in search results, top-10 recommendations, and home page recommendations (this is a precondition hypothesis to all remaining hypotheses under the RQ2).
    \item \textbf{H2.1:} Watching the sequence of misinformation debunking videos after the sequence of misinformation promoting videos will decrease the presence of misinformation in search results, top-10 recommendations, and home page recommendations \emph{in comparison to the end of the promoting sequence}.
    \item \textbf{H2.2:} Watching the sequence of misinformation debunking videos after the sequence of misinformation promoting videos will decrease the presence of misinformation in search results, top-10 recommendations, and home page recommendations \emph{in comparison to the start of the experiment}.
    \item \textbf{H2.3:} The presence of misinformation changes gradually (approximately linearly) with the number of watched videos that promote or debunk misinformation. That is, the misinformation filter bubble effect \emph{increases linearly} with the increasing number of misinformation promoting videos and \emph{decreases linearly} with the increasing number of misinformation debunking videos.
\end{itemize}

In H2.3, we purposefully refer to a gradual linear change as a kind of standard (``baseline'') change, that allows the most straightforward interpretation. This decision is also motivated by the fact that we are not aware of any existing studies on dynamics of misinformation filter bubbles that can be used as a source of assumption for more specific change (like exponential).

\subsection{Metrics and Operationalization of Hypotheses}
\label{sec:methodology:metrics}

To evaluate the hypotheses we use the following metrics:

\begin{description}
    \item [Normalized score (NS).] Drawn directly from the reference study~\cite{Hussein2020}. It quantifies misinformation prevalence in a given list of items (videos), which are labeled as either \emph{promoting} (value 1), \emph{debunking} (value -1) or \emph{neutral} (value 0). It is computed as an average of individual annotations of the items present in a list without considering their order (rank). Its value ranges from the $\left \langle -1, 1  \right \rangle$ interval. Lists populated with mostly debunking content would receive values close to -1, with promoting close to 1, and with balanced or mostly neutral, close to 0. In other words, a score closer to -1 means a better score (less misinformation).
    
    \item [Search result page misinformation score (SERP-MS).] Also drawn directly from the reference study. Unlike the NS metric, it takes into account the rank of the items in a list. It is thus better suited for longer ordered lists. It is computed as~\cite{Hussein2020}: 
    
    $$SERP\mbox{-}MS = \frac{\sum_{r=1}^{n}(x_i*(n-r+1))}{\frac{n*(n+1)}{2}}$$ 
    
    where $x_i$ is annotation value, $r$ search result rank and $n$ number of search results in the list. Its value also ranges from the $\left \langle -1, 1  \right \rangle$ interval with the same interpretation as above. 
    
    \item [Difference to linear (DIFF-TO-LINEAR)] A metric that describes the slope of changes in the normalized score as videos are watched. It compares the actual change in the value of normalized score between a given start and an end video against an expected linear change. We define the final value of the metric as a sum of these differences at each watched video:
    
    $$DIFF\mbox{-}TO\mbox{-}LINEAR = \sum_{i=s}^{e} \left( NS_i - NS_s - \frac{NS_e - NS_s}{e - s}*(i - s) \right)$$ 
    
    where $s$ and $e$ are indices of the start and end videos, $NS_i$ is the normalized score at the $i$-th watched video. If the overall change in the normalized score is positive (i.e., there is an increase in the normalized score on the given interval between the start and end videos) and the metric value is positive, the normalized score changes \textit{faster} than expected. If the metric value is negative, it changes \textit{slower} than expected. In case of a negative overall change in the normalized score (i.e., there is a decrease on the given interval between the start and end videos), the interpretation is reversed. That is, if the metric value is positive, the normalized score changes \textit{slower} than expected and if it is negative, it changes \textit{faster}.
\end{description}

 The \textit{Normalized score (NS)} and \textit{Search result page misinformation score (SERP-MS)} metrics are used to evaluate the following hypotheses: H1.1 (their decrease is expected); H2.0 (their increase is expected); H2.1 (their decrease is expected); and H2.2 (their decrease is expected). The NS metric is used to score shorter lists, which are, in our case, top-10 recommendations and top-10 home page results. On the other hand, SERP-MS metric is used for longer lists of search results. Finally, the DIFF-TO-LINEAR metric is used to evaluate the hypothesis H2.3 (values near 0 are expected).

\subsection{Study scenarios}
\label{sec:methodology:scenarios}

\begin{figure*}[b]
\centering
\includegraphics[width=\textwidth]{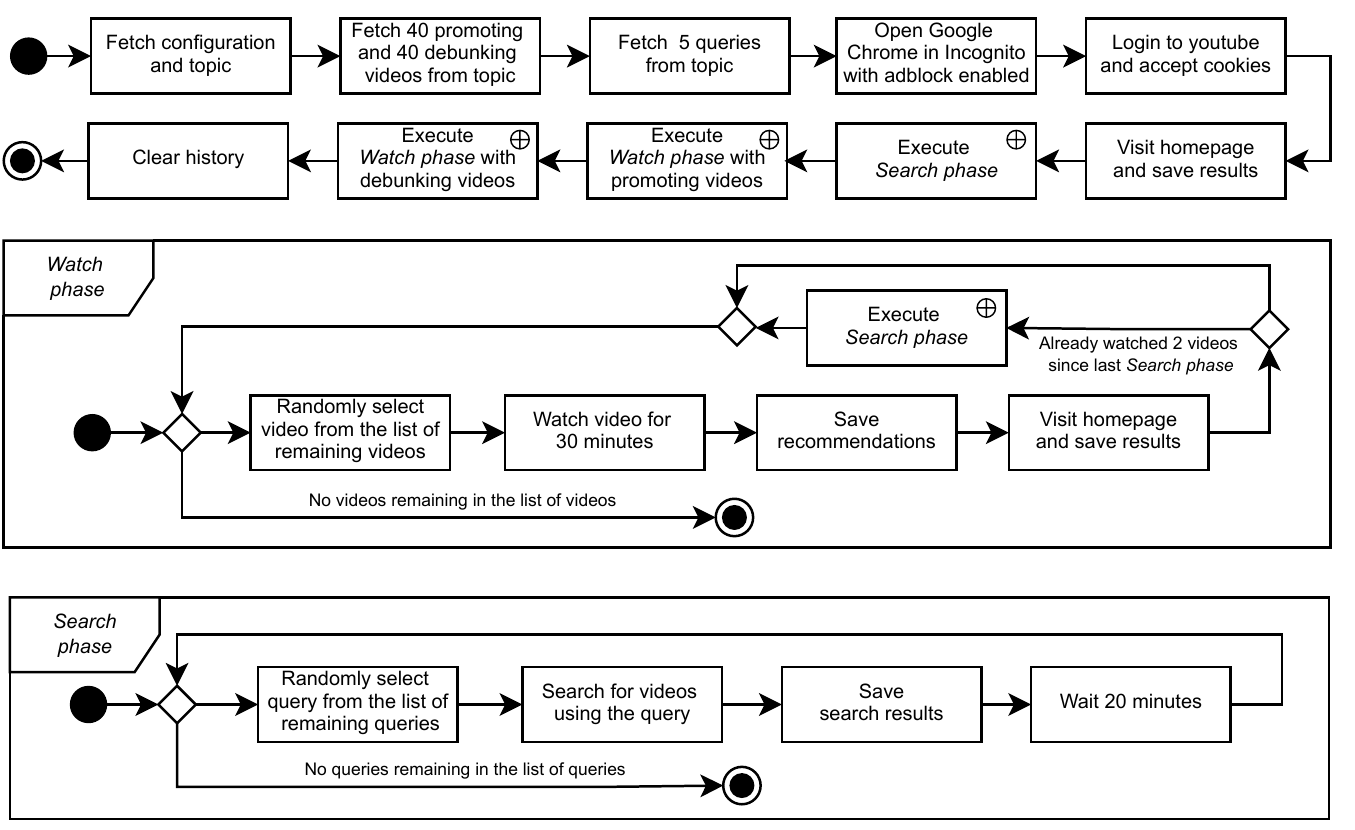}
\caption{The diagram depicting the agent scenario executed during the audit.}
\label{fig:agent-scenario}
\end{figure*}

We let agents interact with YouTube following a \emph{scenario} composed of four phases, as depicted in Algorithm~\ref{algo:scenario} (see also the diagram depicting the agent scenario in the Figure \ref{fig:agent-scenario}).

\begin{algorithm}
	\caption{Creating and bursting a filter bubble}
	\label{algo:scenario}
	\begin{algorithmic}[1]
	    \Procedure{Agent initialization}{}
	        \State Fetch agent configuration (account, controlled parameters - $t_{watch}$, $t_{wait}$, $n_{q}$, $n_{prom}$, $n_{deb}$)
	        \State $\tau$ = Fetch topic from list of possible topics $T$
	        \State $V_{prom}$ = Fetch list of $n_{prom}$ seed videos promoting $\tau$ misinformation
	        \State $V_{deb}$ = Fetch list of $n_{deb}$ seed videos debunking $\tau$ misinformation
	        \State $Q$ = Fetch $n_{q}$ queries from $\tau$
	        \State Initialize browser in incognito mode
	        \State Log in using the given account credentials and accept cookies
	        \State Visit home page and save videos listed there to create a neutral baseline
	        \State Execute \Call{Search}{$Q$} to create a neutral search baseline
	   \EndProcedure
	    
	    \Procedure{Search}{$Q$}
	        \For {query from randomized $Q$}:
	            \State Execute search query and save search results
	            \State Wait $t_{wait}$ minutes
	        \EndFor
	    \EndProcedure
	    
	    \Procedure{Watch}{$V$}
	        \For {video from randomized $V$}:
	            \State Watch video for $t_{watch}$
	            \State Save recommendations found on video's page
	            \State Visit home page and save videos listed there
	            \If {Watched $f_{q}$ videos since the last search}:
	                \State \Call{Search}{$Q$}
	            \EndIf
	        \EndFor
	    \EndProcedure
	    
	    \Procedure{Tear-down}{}
	        \State Clear user history
	    \EndProcedure
	    
	    \State Phase 0: Execute \Call{Agent initialization}{}
	    \State Phase 1: Execute \Call{Watch}{$V_{prom}$}
	    \State Phase 2: Execute \Call{Watch}{$V_{deb}$}
	    \State Phase 3: Execute \Call{Tear-down}{}
	\end{algorithmic}
	\label{algo:basic-scenario}
\end{algorithm}

% Source: https://app.diagrams.net/#G1IM1quXp4PKFuMua91OMqYK1lY4uuyZeg

\emph{Phase 0: Agent initialization.} At the start of a run, the agent fetches its desired configuration, including the YouTube user account and various controlled variables (the variable values are explained further below). Also, the agent fetches $\tau \in T$, a topic with which it will work (e.g., ``9/11''). The agent fetches $V_{prom}$ and $V_{deb}$, which are lists of $n_{prom}=40$ and $n_{deb}=40$ most popular videos promoting, respectively debunking, misinformation within topic $\tau$. Afterwards, it fetches $Q$, a set of $n_{q}=5$ search queries related to the particular $\tau$ (e.g., ``9/11 conspiracy``). The agent configures and opens a browser in incognito mode, visits YouTube, logs in using the given user account, and accepts cookies. Finally, the agent creates a neutral baseline by visiting the home page and saving videos, and performing a search phase. In the \emph{search phase}, the agent randomly iterates through search queries in $Q$, executes each query on YouTube, and saves the search results. To prevent any carry-over effect between search queries, the agent waits for $t_{wait}=20$ minutes after each query.

\emph{Phase 1: Watch promoting videos (create the filter bubble).} For creating a filter bubble effect, the agent randomly iterates through $V_{prom}$ and ``watches'' each video for $t_{watch}=30$ minutes (or less, if the video is shorter). Immediately after watching a video, the agent saves video recommendations on that video's page and visits the YouTube home page, saving video recommendations listed there as well. After every $f_{q}=2$ videos, the agent performs another search phase.

\emph{Phase 2: Watch debunking videos (burst the filter bubble).} The agent follows the same steps as in phase 1. The only difference is the use of $V_{deb}$ instead of $V_{prom}$. 

\emph{Phase 3: Tear-down.} In this phase, the agent clears YouTube history (using Google's ``my activity`` section), making the used user account ready for the next run.

For each selected topic, we run the scenario 10 times (in parallel). This way, we were able to deal with recommendation noise present at the platform. In order to run our experiments multiple times, we used the \emph{reset} (delete all history) button provided by Google instead of creating a new user profile for each run. %We delete the whole history in one go, not only from the YouTube platform but also from all other Google products, such as Google Ads and Google Search.
Before deciding to use the \emph{reset} button in our study, we first performed a short verification study to see whether using this button really deletes the whole history and resets the personalization on YouTube. We randomly selected few topics, from which we manually watched few videos (5 for each). Then, we used the reset button and evaluated the difference between videos appearing on the YouTube home page, recommendations, and search. We found no carry-over effects.

\subsubsection{Controlled variables and parameter setup pre-study}

% The comments here are just for reference

% $n_q$ - number of search queries
% $Q$ - set of queries
% $\tau$ - topic selected for the run
% $T$ - list of topics selected for experiments

% $V$ - list of all videos
% $V_{prom}$ - list of promoting videos
% $V_{deb}$ - list of debunking videos

% $n_{prom}$ - number of promoting videos
% $n_{deb}$ - number of debunking videos

% $t_{wait}$ - waiting time after querying
% $t_{watch}$ - watching time for one video

% $f_{q}$ - number of videos to watch before using search queries

In the study, we control several sources of potential confounds:

\begin{itemize}
    \item \textit{Geolocation} from which the agents access the YouTube service and other \textit{user characteristics} of the agents (name, date of birth, gender). 
    \item \textit{Time}, i.e., all experiments are done in shortest time frame possible to minimize the influence of content newly appearing on YouTube. Risks of changes in YouTube's recommendation algorithms during the audit are also minimized this way.
    \item \textit{Technical setup of agents} uses always the same configuration (e.g., operating system, browser) as described below.
\end{itemize}

For \emph{geolocation}, we use N.~Virginia to allow for better comparison with the reference study. The date of birth for all accounts was arbitrarily set to 6.6.1990 to represent a person roughly 30 years old. The gender was set as ``rather not say'' to prevent any personalization based on gender. The names chosen for the accounts were composed randomly of the most common surnames and unisex given names used in the US.

There were also \emph{process parameters} that we needed to keep constant. These include: 

\begin{enumerate}
    \item $n_{prom}=40$ and $n_{deb}=40$ representing the number of seed videos used in the promoting and the debunking phase respectively,
    \item $t_{watch}=30$ representing the maximum watching time in minutes for every video,
    \item $n_{q}=5$ representing the number of queries used,
    \item $t_{wait}=20$ representing the wait time in minutes between query yields, and
    \item $f_{q}=2$ representing the number of videos to watch between search phases.
\end{enumerate}

Values of the \emph{process parameters} greatly influence the total running time and results of the study. Yet, determining them was not straightforward given many unknown properties of the environment (first and foremost YouTube's algorithms). For example, prior to the study, it was unclear how often we need to probe for changes in recommendations and search result personalizations to answer our research questions.

Therefore, \emph{we ran a pre-study in which we determined the best parameter setup}. We used two metrics to determine the type of change (i.e., change in videos, or change in order of videos) between lists of returned videos: 1) the \emph{overlap} of lists of recommended videos, a simple metric that disregards order of videos; and 2) the \emph{Levenshtein distance} between ordered results that takes change in order, but not the exact position, into consideration. Measuring these metrics across watched videos and different runs, we determined to run 10 individual agents for each topic, as we observed instability between repeated runs (e.g., the same configuration yielded \(\sim70\%\) of the same recommended videos). For the $n_{prom}$ and $n_{deb}$ parameters, we observed that in some cases, a filter bubble effect (a change in diversity of the returned videos) could be detected after 20 watched videos. Yet in others, it was after 30 or more. Due to this inconsistency, we opted to watch 40 videos for a phase to guarantee that the full potential of misinformation filter bubble effect is developed and observed. To determine the optimal value of $t_{watch}$, we first calculated the average running time of our seed videos. Most of the videos (\(\sim85\%\)) had a running time of about 30 minutes or shorter, so 30 minutes became the baseline value. In addition, we compared the results obtained by watching only 30 minutes with results from watching the whole video regardless of its length, but found no apparent differences. 

To determine the number of queries $n_{q}$ and periodicity of searches $f_{q}$, we ran the scenario with all seed queries introduced by the reference study and used them after every seed video. We observed that the difference in search results between successive seed videos was not significant. As the choice of search queries and the frequency of their use greatly prolonged the overall running time of the agents, we opted to run the search phase after every second video. In addition, we opted to use only 5 queries per topic.

The only parameter not set by a pre-study is $t_{wait}$, which we set to 20 minutes based on previous studies. These found that the carry-over effect (which we wanted to avoid) is visible for 11 minutes after the search~\cite{Hannak2013, Hussein2020}.

\subsubsection{Technical setup}

The implementation of the agents is available in the accompanying GitHub repository. They were implemented using the Selenium library in Python. We used Google Chrome browser version 88 with chromedriver version 88.0.4324.96. We used the incognito mode to remove any possibility of history or cookies being carried over across different sessions. In addition, we used uBlock Origin to prevent noise from watching ads in our experiments. Each agent was implemented using Python 3.8.7 Dockerfile based on Debian 10.7. All the agents were placed at an AWS server running in N.~Virginia.

\subsection{Seed Data}
\label{sec:methodology:seed}

We used 5 topics in our study (same as the reference study): 1) \emph{9/11 conspiracies} claiming that authorities either knew about (or orchestrated) the attack, or that the fall of the twin towers was a result of a controlled demolition, 2) \emph{moon landing conspiracies} claiming the landing was staged by NASA and in reality did not happen, 3) \emph{chemtrails conspiracy} claiming that the trails behind aircrafts are purposefully composed of dangerous chemicals, 4) \emph{flat earth conspiracy} claiming that we are being lied to about the spherical nature of Earth, and 5) \emph{vaccines conspiracy} claiming that vaccines are harmful, causing various range of diseases, such as autism. The narratives associated with the topics are \emph{popular} (persistently discussed), while at the same time, \emph{demonstrably false}, as determined by the reference study~\cite{Hussein2020}.

For each topic, the experiment required two sets of seed videos. The \emph{promoting} set, used to construct a misinformation filter bubble (its videos have a promoting stance towards the conspiratorial narrative or present misinformation). And the \emph{debunking} set, aimed to burst the bubble (and containing videos disproving the conspiratorial narratives).

As a basis for our seed data sets we used data already published in the reference study, which the authors either used as seed data, or collected and annotated. To make sure we use adequate seed data, we re-annotated all of them.

The number of seed videos collected this way was insufficient for some topics (we required twice as many seed videos as the reference study). To collect more, we used an extended version of the seed video identification methodology of the reference study. Following is the list of approaches we used (in a descending order of priority): YouTube search, other search engines (Google search, Bing video search, Yahoo video search), YouTube channel references, recommendations, YouTube home page, and known misinformation websites. To minimize any biases, we used a maximum of 3 videos from the same channel. 

As for search queries, we required fewer of them than the reference study. We selected a subset based on their popularity on YouTube. Some examples of the used queries are: ``\emph{9/11 conspiracy}'', ``\emph{Chemtrails}'', ``\emph{flat earth proof}'', ``\emph{anti vaccination}'', ``\emph{moon landing fake}''.

The full list of seed videos and used search queries is available in the accompanying GitHub repository.

\subsection{Data collection and annotation}
\label{sec:methodology:annotation}

Agents collect videos from three main components on YouTube: 1) \emph{recommendations} appearing next to the watched videos, 2) \emph{home page} videos, and 3) \emph{search results}. In case of recommendations, we collect first 20 videos that YouTube displays next to a currently watched video (in rare cases, less than 20 videos are recommended).  Collecting top-\(N\) videos is in line with previous works and sources from the fact that users seldom visit recommendations below. For home page videos and search results, we collect all videos appearing with the given resolution (2560x1440px), but no less than 20. In case when less than 20 videos appear, the agent scrolled further down on the page to load more videos.

For each video encountered, the agent collects the following metadata: 1) \emph{YouTube video ID}, 2) \emph{position} of the video in the list, and 3) \emph{presence of a warning/clarification message} that appears with problematic topics such as COVID-19. Other metadata, such as video \emph{title}, \emph{channel}, or \emph{description} are collected using the YouTube API.

To annotate the collected videos for the presence of misinformation, we used an extended version of the methodology proposed in the reference study. Each video was viewed and annotated by the authors of this study using a code ranging from -1 to 10 as follows: 
\begin{itemize}
    \item \textit{Code -1}, i.e., \emph{debunking}, when the narrative of a video provides arguments against the misinformation related to the particular topic (such as \emph{``The Side Effects of Vaccines - How High is the Risk?''}).
    \item \textit{Code 0}, i.e., \emph{neutral}, when the narrative discusses the related misinformation but does not present a stance towards it (such as \emph{``Flat Earthers vs Scientists: Can We Trust Science? | Middle Ground''}).
    \item \textit{Code 1}, i.e., \emph{promoting}, when the narrative promotes the related misinformation (such as \emph{``MIND BLOWING CONSPIRACY THEORIES''}).
    \item \textit{Codes 2, 3, and 4} have the same meaning as codes -1, 0, and 1, but are used in cases when they discuss misinformation not related to the topic of the run (e.g., video dealing with climate crisis misinformation encountered during a flat earth audit).
    \item \textit{Code 5} is applied to videos that do not contain any misinformation views (such as \emph{``Gordon's Guide To Bacon''}). This includes completely unrelated videos (e.g., music or reality show videos), but also videos that are related to the general audit topic, but not misinformation (e.g., original news coverage of 9/11 events).
    \item \textit{Code 6} is assigned in rare cases of videos that are not in English and do not provide English subtitles.
    \item \textit{Code 7} is assigned in equally rare cases when the narrative of the video cannot be determined with enough confidence.
    \item \textit{Code 8} is reserved for videos removed from YouTube (before they are annotated).
    \item \textit{Codes 9 and 10} present an extension of the approach used in the reference study. They are used to denote videos that specifically mention misinformation but rather than debunk them, they mock them (9 for related misinformation, 10 for unrelated misinformation, for example \emph{``The Most Deluded Flat Earther in Existence!''}). Mocking videos are a distinct (and often popular) category, which we wanted to investigate separately (however, for the purposes of analysis, they are treated as debunking videos).
\end{itemize}

Similar to the reference study, we map all the codes assigned by the annotators to one of the three stance values: -1 (debunking), 1 (promoting), and 0 (neutral), which are used to compute the metrics and evaluate our hypotheses. Codes -1, 2, 9, and 10 are mapped to -1. Codes 1 and 4 are mapped to 1. Codes 0, 3, and 5 are mapped to 0. Videos coded with 6, 7, or 8 are discarded from the evaluation, since they cannot be reliable mapped to any of the stance values.

To determine how many annotators are needed per video, we first re-annotated the seed videos released by the reference study. Each seed video was annotated by at least two authors. To measure an inter-rater reliability, we evaluated consistency between: 1) our annotators, who produced re-annotated labels, achieving Cohen's kappa value of 0.815; and 2) our annotators and authors of the reference study, achieving Cohen's kappa value of 0.688. We attribute the difference between these values to a possible smaller divergence of annotation methodology execution of our and reference study annotators. We identified characteristics of edge cases and counseled how to resolve them with all our annotators. Following the re-annotation and the findings from it, when annotating our collected videos, we assign only one annotator per collected video. Annotators were instructed to report edge cases, i.e., videos hard to label with enough confidence for some reason. Such videos were encountered approximately every 10-20 annotations, and were always reviewed by another annotator (i.e., providing the ``second opinion'') and optionally discussed until a consensus was reached between the annotators.

For the purpose of this study and to evaluate our hypotheses, we manually annotated the following subset of collected videos:
\begin{itemize}
    \item All recorded \emph{search results}.
    \item Videos recommended for the first 2 seed videos at the start of the run and the last 2 seed videos of both phases (resulting in 6 sets of annotated videos per topic). This selection was a compromise between representativeness, correspondence to the reference study, and our capacities.
    \item We did \emph{not} annotate the \emph{home page videos} for the purpose of this study. These videos were the most numerous, the most heterogeneous, and with little overlap across bots and seed videos.
\end{itemize}

For the remaining videos from top-10 recommendations and home page results, which we did not annotate manually, we employed a machine learning model trained on the manual annotations to predict their labels as discussed next.

\subsection{Trained machine learning models for automated prediction of annotations}
\label{sec:methodology:ml-training}

We opted for automatic annotation of the remaining videos by the means of machine learning due to their large number (17,405 unique videos overall, cf. Section~\ref{sec:results}) which would not be feasible to annotate manually. We experimented with two state-of-the-art models for classification of YouTube videos used in similar misinformation detection-related tasks that were presented in the related works---models by Hou et al.~\cite{Hou2019} and Papadamou et al.~\cite{Papadamou2020}.

\subsubsection{Model by Hou et al.~\cite{Hou2019} (Hou's model)}

The authors presented an SVM model trained to classify prostate cancer videos as misinformative or trustworthy based on a set of viewer engagement features (e.g., number of views, number of thumbs ups, number of comments), linguistic features (e.g., n-grams and syntax based features, readability and lexical richness features derived from the video transcripts), and raw acoustic features. We implemented this model using standard ML toolkits (nltk, sklearn) and trained it using our annotated dataset. We omitted using acoustic features in our training since we did not collect them in our dataset. We also experimented with an XGBoost version of this model, which used the same set of input features. We published our implementation of the model (both SVM and XGBoost versions) in the GitHub repository accompanying this paper.

\subsubsection{Model by Papadamou et al.~\cite{Papadamou2020} (Papadamou's model)}

A deep learning model was used to classify YouTube videos related to common conspiracy theory topics as pseudoscientific or scientific. The proposed classifier takes four feature types as input: snippet (video title and description), video tags (defined by the creator of the video), transcript (subtitles uploaded by the creator of the video or auto-generated by YouTube), and top-200 video comments. It then uses fastText (fine-tuned to the inputs) to generate vector representations (embeddings) for each of the textual inputs. Resulting features are flattened into a single vector and processed by a 4-layer, fully-connected neural network (comprising 256, 128, 64, and 32 units with ReLU activation). Regularization using dropout ($d=0.5$) is applied at each fully-connected layer. Finally, the output is passed to a 2-unit layer with softmax activation. There is a threshold for predicting the ``pseudoscientific'' class that requires the classification probability to be 0.7 or higher for it to be used. The classifier is implemented using Keras and Tensorflow. Due to class imbalance (Papadamou's dataset contained 1,325 pseudoscience and 4,409 other videos), oversampling is applied during training to produce the same number of training samples for both classes. We made use of the source code provided by the authors of the paper\footnote{\url{https://github.com/kostantinos-papadamou/pseudoscience-paper}}. However, we did not use video tags as input features as we lacked them in our dataset. Similarly to the authors' original work, we also experimented with a BERT-based version of the model, which used a pre-trained BERT model\footnote{\url{https://tfhub.dev/tensorflow/bert_en_uncased_L-12_H-768_A-12/4}} instead of fastText to compute embeddings of the textual inputs, with the rest of the neural network's architecture remaining the same.

\subsubsection{Classification tasks}

Both models were originally applied for binary classification tasks and classified videos as misinformative/trustworthy in case of Hou's model and pseudoscientific/scientific in case of Papadamou's model. Since our data was annotated with multiple labels that were normalized into three classes (promoting, debunking, neutral), we had to make a decision on how to handle the ``neutral'' class not considered in the original models. We experimented with the following variations of classes in our cross-validation of the models:

\begin{enumerate}
    \item \textit{Binary without neutral}: only promoting (class 1) and debunking (class 2), discarding the neutral videos.
    \item \textit{Binary with neutral}: promoting (class 1) and debunking together with neutral (class 2).
    \item \textit{3 classes}: promoting (class 1), debunking (class 2), and neutral (class 3).
\end{enumerate}

\subsubsection{Performance of the models}

We trained the models using a subset of all manually labeled data (a combination of the seed data and the videos encountered during data collection) for which we could retrieve all necessary information from the YouTube API (such as transcript and other metadata). It consisted of 2,622 labeled videos (405 promoting, 758 debunking, and 1,459 neutral videos). We evaluated the models using 5-fold cross-validation in case of Hou's model and 10-fold cross-validation in case of Papadamou's model to reflect the evaluation in their respective papers. Table~\ref{tab:cv-metrics-2-classes} shows classification metrics comparing the models' performance reported in the original papers with their performance on our data evaluated on a binary setup without the neutral class and with it. Table~\ref{tab:cv-metrics-3-classes} compares the models' performance on a 3 classes setup. It also includes evaluation of an XGBoost version of the Hou's model and a BERT-based version of the Papadamou's model.

Hou's model showed performance similar to that reported in the paper when applied to the binary classification task with only the promoting and debunking classes. On the other hand, the performance (on the promoting class) decreased when we incorporated neutral videos into a ``debunking + neutral'' class (cf. Table~\ref{tab:cv-metrics-2-classes}). The low precision (0.42) on promoting class shows that the model does not have predictive power to distinguish these classes. Applying the model to classification of all three classes showed relatively weak performance as well (cf. Table~\ref{tab:cv-metrics-3-classes}). When comparing the results of the SVM model originally used in Hou et al.'s work~\cite{Hou2019} to the results of the XGBoost model, we can see that their performance is similar with SVM having slightly better performance on the promoting class.

Papadamou's model achieved better performance when applied to binary classification with promoting and debunking videos only and also outperformed the metrics reported in the original paper (cf. Table~\ref{tab:cv-metrics-2-classes})---we attribute this improvement to the quality of our data which was annotated by experts instead of crowd-sourcing annotators who were employed by Papadamou et al. It also retained a good performance (0.71 F1-score on the promoting class) when neutral videos were added into the ``debunking + neutral'' class. Therefore, we decided to adapt this model for classification of all three classes: promoting, debunking, and neutral. In this task, the model achieved a slightly lower F1-score (0.66) caused by lower recall (0.65 compared to 0.76) on the promoting class (with 405 samples); cf. Table~\ref{tab:cv-metrics-3-classes}. On the debunking class (with 758 samples), the model achieved precision 0.79 and recall 0.74. On the neutral class (with 1,459 samples), precision was 0.86 and recall 0.91. The results of the BERT-based version of the model were worse than the original Papadamou's model trained on our data and it was even outperformed by both SVM and XGBoost versions of the Hou's model. This is in line with the results reported in~\cite{Papadamou2020}, where the BERT-based model performed worse than the fastText version and had mixed results when compared to SVM and Random Forest. Table~\ref{tab:cv-confusion-matrix} shows a confusion matrix for the best-performing Papadamou's model on all three classes.

\subsubsection{Implications for the rest of the paper}

Seeing that Hou's model was struggling with the neutral class, we opted for Papadamou's model for the use in this paper. We further decided to take advantage of the model trained for the 3-class classification task as that enables deeper analyses and retains a satisfactory performance.

\begin{table}
\caption{
    Comparison of the classification metrics of the evaluated models as reported in their original papers (training: ``Paper'') or cross-validated on our data (training: ``Our data''). We experimented with two different class setups to train the models on (classes: ``Binary without neutral'' and ``Binary with neutral''). Precision, recall, and F1-score are reported both on the promoting (prom.) class (misinformative in the paper by Hou et al.~\cite{Hou2019}, not reported by Papadamou et al.~\cite{Papadamou2020}), as well as their weighted (weigh.) average across classes.
}
\label{tab:cv-metrics-2-classes}
\centering

\begin{tabular}{r|cc|cc|cc}
\toprule
Training            & \multicolumn{2}{c|}{Paper}  & \multicolumn{4}{c}{Our data}  \\
Classes & \multicolumn{2}{c|}{Binary} & \multicolumn{2}{c|}{Binary w/o neutral} & \multicolumn{2}{c}{Binary w neutral} \\
Model               & Hou & Papad. & Hou & Papad. & Hou & Papad.  \\
\midrule
Precision prom. & 0.77 &   -  & 0.70 & 0.82 & 0.42 & 0.68 \\
Recall prom.    & 0.73 &   -  & 0.65 & 0.85 & 0.63 & 0.76 \\
F1-score prom.  & 0.72 &   -  & 0.68 & 0.83 & 0.51 & 0.71 \\
\hline
Precision weigh.& 0.78 & 0.77 & 0.78 & 0.91 & 0.85 & 0.93 \\
Recall weigh.   & 0.74 & 0.79 & 0.78 & 0.91 & 0.81 & 0.93 \\
F1-score weigh. & 0.74 & 0.74 & 0.78 & 0.91 & 0.82 & 0.93 \\
\hline
Accuracy        & 0.74 & 0.79 & 0.78 & 0.91 & 0.81 & 0.93 \\
\bottomrule
\end{tabular}
\end{table}

\begin{table}
\caption{
    Comparison of the classification metrics of the evaluated models cross-validated on our data with the ``3 classes'' setup. Precision, recall, and F1-score are reported both on the promoting (prom.) class as well as their weighted (weigh.) average across classes. We compare the original models proposed by Hou et al.~\cite{Hou2019} and Papadamou et al.~\cite{Papadamou2020} to their XGBoost and BERT-based variants respectively. For the data analysis presented further in this paper, we make use of the best-performing model, which is reported in the rightmost column of this table (text in italics)---model from Papadamou et al. classifying videos into 3 classes (promoting, debunking, and neutral).
}
\label{tab:cv-metrics-3-classes}
\centering

\begin{tabular}{r|cc|cc}
\toprule
Classes & \multicolumn{4}{c}{3 classes}  \\
Model   & \multicolumn{2}{c|}{Hou} & \multicolumn{2}{c}{Papadamou} \\
Variant & XGBoost & SVM (orig.) & BERT & fastText (orig.)\\
\midrule
Precision prom. & 0.49 & 0.59 & 0.61 & \emph{0.68} \\
Recall prom.    & 0.48 & 0.46 & 0.38 & \emph{0.65} \\
F1-score prom.  & 0.49 & 0.51 & 0.47 & \emph{0.66} \\
\hline
Precision weigh.& 0.73 & 0.73 & 0.69 & \emph{0.81} \\
Recall weigh.   & 0.74 & 0.74 & 0.69 & \emph{0.82} \\
F1-score weigh. & 0.73 & 0.73 & 0.68 & \emph{0.82} \\
\hline
Accuracy        & 0.74 & 0.74 & 0.70 & \emph{0.82} \\
\bottomrule
\end{tabular}
\end{table}

\begin{table}
\caption{
    Confusion matrix from cross-validation of the model by Papadamou et al.~\cite{Papadamou2020} trained on our data for classification into three classes. There is a significant class imbalance with the neutral class being the most prominent. Oversampling was used in training to address this problem.
}
\label{tab:cv-confusion-matrix}
\centering

\begin{tabular}{r|ccc}
\toprule
          & promoting (predicted) & neutral (predicted)    & debunking (predicted) \\ \midrule
promoting (actual) & 262 (65\%) & 75 (19\%)   & 68 (17\%)  \\
neutral (actual)   & 55 (4\%) & 1,324 (91\%) & 80 (5\%)  \\
debunking (actual) & 66 (9\%)  & 132 (17\%)   & 560 (74\%) \\ \bottomrule
\end{tabular}
\end{table}

\subsection{Ethics assessment}
\label{sec:methodology:ethics}

To consider the various ethical issues regarding the research of misinformative content,  we participated in a series of ethics workshops facilitated by experts on AI and data ethics from among the co-authors of this paper who were not part of the technical team. The workshops were aimed at exploring questions related to data ethics~\cite{Tranberg2020Dataethics} and AI ethics issues~\cite{ALTAI} within our audit and its impact on stakeholders in the light of the principles of Responsible Research and Innovation~\cite{rri2012}. The most affected stakeholder groups were platform users, annotators, content creators, and researchers. We devised different engagement strategies and specific action steps for every stakeholder group. Our main task was to devise countermeasures to the most prominent risks that could emerge for these stakeholder groups.

First, we were concerned about the risk of unjustified flagging of the content as misinformation and their creators as conspirators. To minimize this risk, we decided to report hesitations in the manual annotation process. These hesitations were consequently back-checked by other annotators and independently validated until the consensus was reached. After the deployment of the automated machine learning pipeline, this concern was accompanied by the risk that we might over-rely on automated annotation based on previous results and lose caution in the human annotating process. Therefore, we proposed to put in place processes to check the performance of the model and to detect its potential deterioration to ensure that human oversight is in place for video recommendations and home page results. These include the proposal for manual annotation of a new sample of data (in case we decided to use the trained model on a newly collected data, e.g., collected later in time or for different topics, which was, however, not yet the case) and checking properties of the new data, e.g., distributions compared to the training and validation data.

One of our main concerns was also not to harm or delude other users of the online platform. To avoid a disproportional boost of the misinformation content by our activity, we selected videos with at least 1000 views and warned the annotators not to watch videos online more than once, or in case of back-checks, twice. After each round, we reset the user account and deleted the watch history. To minimize the risk of misuse of data about users and their alleged inclination to possibly misinformative content, which can be presented, e.g., in video comments, we decided not to publish automatically annotated data and store this data at servers with appropriate safeguards in place with access only to approved researchers.

Other concerns were connected to the deterioration of the well-being of human annotators. Specifically, their decision-making abilities would be negatively affected after a long annotation process. The ethics experts from among the co-authors of this paper, who did not participate on the annotation process, proposed daily routines for the annotation, including breaks during the process, and advised to monitor any changes in the annotator's beliefs. The annotators also underwent a survey on their tendency to believe in conspiracy theories\footnote{\url{https://openpsychometrics.org/tests/GCBS/}} and none of them showed such tendency at the end of the study.

We also wanted to avoid the situation of being too general in our conclusions considering the dynamics of filter bubble creation and bubble bursting for various groups of users. Together with ethics experts we identified possible limitations of our study that were not only tied to the amount of topics that we have investigated or the set of agent interactions with the platform, but also to the fact that the agent's user account was tied only to specific settings (geolocation, date of birth, gender or name) despite the fact that we have no evidence of their impact on the filter bubble creation dynamics.
Yet, we note that there still remain the risks that the conclusions drawn from the results of our study might not be fully representative of other groups of users. To validate our results on a more diverse set of users is a future challenge together with the need for the deployment of tools to identify potential sources of systematic discrimination or other unwanted biases in data or the model. 

We were also aware that the risk of possible unjust biases and discrimination presented in the model is tightly connected to the problem of transparency and explainability of the proposed model for automated annotation. This problem still poses an open question for future research. We have decided to publish the source code and datasets excluding automatically annotated data to support any future research in this area and maintain the privacy of users at the same time. In addition, to avoid any possible legal issues, we made public only YouTube IDs of seed and encountered videos without any metadata, such as titles, description or transcripts, which might be protected by copyrights. However, these can be downloaded using YouTube API (if the video is still available on the platform).

\subsection{A note on comparability with the reference study by Hussein et al.}
\label{comparison-explained}

In order to be able to draw comparisons, we kept the methodology of our study as compatible as possible with the previous study by Hussein et al.~\cite{Hussein2020}. We shared the general approach of prompting YouTube with implicit feedback: both studies used similar scenarios of watching a series of misinformation promoting videos and recording search results and recommended videos. We re-used the topics, a subset (for scaling reasons) of search queries, and all available seed videos (complementing the rest by using a similar approach as the reference study). Moreover, both studies used the same coding scheme, metrics, sleep times, and annotated a similar number of videos. 

We should also note differences between the studies, which mainly source from different original motivations for our study. For instance, no significant effects of demographics and geolocation of the agents were found in the reference study, so we only controlled these. In Hussein's experiments, all videos were first ``watched'' and only then all search queries were fired. In our study, we fired all queries after watching \emph{every 2nd} video (with the motivation to get data from the entire run, not just the start and end moment). The reference study created genuine 150 accounts on YouTube, while we used fewer accounts and took advantage of the browsing history reset option. In some aspects, our study had a larger scale: we executed 10 runs for each topic instead of one (to reduce possible noise) and used twice as many seed videos (to make sure that filter bubble effects develop). There were also technical differences between the setups, as we used our own implementation of agents (e.g., different browser, ad-blocking software).

Given the methodological alignment (and despite the differences), we are confident to directly compare some of the outcomes of both studies, namely quantity of misinformative content appearing at the end of the promoting phases.
\section{Results and findings}
\label{sec:results}

Following the study design, we executed the study between March 2nd and March 31st, 2021. Together, we executed 50 bot runs (10 for each topic). On average, runs for a single topic took 5 days (bots for a topic ran in parallel). The bots watched 3,951 videos, executed 10,075 queries, and visited home page 3,990 times. For each of these, we recorded the recommendations and results provided by YouTube as shown in Table~\ref{tab:collected-data}. Overall, we recorded 17,405 unique videos originating from 6,342 channels.

\begin{table}
\caption{
    Overview of the data collected by bots during the study execution.
}
\label{tab:collected-data}
\centering

\begin{tabular}{r|cc}
\toprule
                  & \#all encountered videos & \#unique videos \\ \midrule
Recommendations   & 78,763 & 8,526  \\
Search results    & 201,404 & 942  \\
Home page results & 116,479 & 9,977 \\ \hline
Total             & 396,646 & 17,405 \\ \bottomrule
\end{tabular}
\end{table}

Using the selection strategy and annotation scheme described in Section~\ref{sec:methodology:annotation}, five annotators annotated 2,914 unique videos (covering 255,844 appearances). The distribution of labels is shown in Table~\ref{tab:labels-distribtution}. Promoting videos constituted 8\% of annotated search results, and 7\% of annotated top-10 recommendations. Debunking videos made up 27\% of annotated search results, and 19\% of annotated top-10 recommendations. Using the trained machine learning model, we retrieved labels for additional 13,801 videos from top-10 recommendations and home page results. Their distribution is also shown in Table~\ref{tab:labels-distribtution}. Promoting videos constituted 3\% of predicted top-10 recommendations, and 4\% of predicted home page results. Debunking videos made up 56\% of predicted top-10 recommendations, and 44\% of predicted home page results. Compared to the manually annotated data, we see lower percentages of promoting videos, but higher percentages of debunking videos.

\begin{table}
\caption{
    Distribution of manual and predicted labels. Promoting and debunking videos include those related as well as unrelated to respective topics. Debunking videos also include those manually annotated as mocking videos. In further analyses presented in the paper as well as in the classification model, we do not distinguish between neutral videos and videos labeled as not about misinformation (both are regarded as neutral). The other videos include videos that were manually labeled as unknown, non-English or removed.
}
\label{tab:labels-distribtution}
\centering

\begin{tabular}{r|cc}
\toprule
                  & Manual labels & Predicted labels \\ \midrule
Promoting   & 244 & 478 \\
Debunking   & 628 & 6,535 \\
Neutral     & 184 & \multirow{2}{*}{6,788}  \\
Not about misinformation & 1,829 & \\
Other       & 29 & - \\ \hline
Total       & 2,914 & 13,801\\ \bottomrule
\end{tabular}
\end{table}

Table~\ref{tab:data-descriptive} shows basic descriptive statistics of the collected data with respect to the length of videos and how many times they appeared during data collection. Both distributions are power-law distributions with very long tails. When looking at more frequent videos (appearing 10 or more times), we observe that even though they represent only around 17\% of the collected data, they comprise 26\% of all promoting videos, i.e., they are 1.7 times more likely to be promoting misinformation than the less frequent videos. We also collected the popularity of the videos (number of views, comments, likes, etc.), but due to many missing values (in about 40\% of all encountered videos), we do not report statistics related to the popularity.

\begin{table}
\caption{
    Descriptive statistics of the collected data with respect to the length of videos (\#minutes) and how many times they appeared during data collection (\#encounters). The statistics are computed for a subset of 15,837 videos (out of all 17,405 encountered videos), for which we were able to obtain metadata. We collected videos' metadata using YouTube API some time after the data collection itself. This meant that we were not able to get metadata for all encountered videos, e.g., in cases when the videos were removed by the authors or by the platform.
}
\label{tab:data-descriptive}
\centering

\begin{tabular}{r|ccccccc}
\toprule
             & Mean & Std. & 25\% & 50\% & 75\% & 90\% & Max. \\ \midrule
\#minutes    & 45.42 & 108.87 & 7.73 & 14.73 & 39.72 & 92.69 & 1439.88 \\
\#encounters & 20.4 & 112.92 & 1 & 2 & 5 & 21 & 2695 \\ \bottomrule
\end{tabular}
\end{table}

We report the results according to research questions and hypotheses defined in Section~\ref{sec:methodology:rqs}. We use manually annotated data to answer all research questions and to test our hypotheses except for those related to home page results and the hypothesis H2.3 concerned with the slope of change in the proportion of misinformation; in those cases we also use the automatically labeled data to complement the manual labels.
SERP-MS score metrics are reported for search results and mean normalized scores for recommendations and home page results.
Since the metrics are not normally distributed with some samples of unequal sizes, we make use of non-parametric statistical tests.
Pairwise tests are performed using two-sided Mann-Whitney U test.
In cases where multiple comparisons by topics are performed, Bonferroni correction is applied on the significance level (in that case \(\alpha=0.05\) is divided by number of topics \(n_T=5\), resulting in \(\alpha=0.01\)).

\subsection{RQ1: Has YouTube's personalization behavior changed since the reference study?}

%%% Overall results across topics
Regarding H1.1, we overall see a small change in the mean SERP-MS score across the same search queries in our and reference data: mean SERP-MS worsened from -0.46 (std 0.42) in reference data to -0.42 mean (std 0.3) in our data.
However, the distributions are not statistically significantly different (n.s.d.).
There is a similar small change towards the promoting spectrum in up-next (first result in recommendation list) and top-5 recommendations (following 5 recommendations).
We compared the up-next and top-5 recommendations together (as top-6 recommendations) using last 10 watched promoting videos in reference watch experiments and last two watched videos in our promoting phase.
We see mean normalized score worsened from -0.07 (std 0.27) in reference data to -0.04 (std 0.31) in our data.
These distributions are also not significantly different (U=45781.5, n.s.d.).

%%% Results within topics
More considerable shifts in the data can be observed when looking at individual topics.
Table~\ref{tab:hussein-comparison-search} shows a comparison of SERP-MS scores for top-10 search results between our and reference data.
Improvement can be seen within certain queries for the chemtrails conspiracy that show a large decrease in the number of promoting videos.
The reference study reported that this topic receives significantly more misinformative search results compared to all other topics.
In our experiments, their proportion was lower than in the 9/11 conspiracy.
On the other hand, search results for flat earth conspiracy worsened.
Queries such as ``flat earth british'' resulted in more promoting videos, likely due to new content on channels with similar names.
Within the anti-vaccination topic, there is an increase in neutral videos (from 12\% to 35\%) and thus a drop in debunking videos (from 85\% to 61\%).
This may relate to new content regarding COVID-19.

Table~\ref{tab:hussein-comparison-recommendations} shows a comparison of normalized scores for up-next and top-5 recommendations.
Only the moon landing and anti-vaccination topics come from statistically significantly different distributions.
Similar to search results, recommendations for the 9/11 and anti-vaccination conspiracy topics worsened.
There were more promoting videos on the 9/11 topic (27\% instead of 18\%).
In the anti-vaccination topic, we observed a drop in debunking videos (from 29\% to 9\%) and a subsequent increase in neutral (from 70\% to 78\%) and promoting videos (from 1\% to 8\%).
The change within the anti-vaccination controversy is even more pronounced when looking at up-next recommendations separately.
Within up-next, the proportion of debunking videos drops from 77\% to 19\%, neutral videos increase from 22\% to 70\%, and promoting increase from 1 to 11\%.
On the other hand, in the moon landing topic, we see much more debunking video recommendations---40\% instead of 23\% in reference data.

%%% Interpretation of results
These results bring up a need to distinguish between \emph{endogenous} (e.g., changes in algorithms, policy decisions made by platforms to hide certain content) and \emph{exogenous} factors (e.g., changes in content, external events, behavior of content creators) as discussed by Metaxa et al.~\cite{Metaxa2019}.
Our observations show that search results and recommendations were in part influenced by exogenous changes in content on YouTube.
Within the chemtrails conspiracy, we observed results related to a (then) new song by Lana del Rey that mentions ``Chemtrails'' in its name.
Search results and recommendations in the anti-vaccination topic seem to be influenced by COVID-19.
Flat earth conspiracy videos were influenced by an increased amount of activity within a single conspiratorial channel.

\begin{table*}
\caption{
    Comparison of SERP-MS scores for top-10 search results with data from the reference study.
    The scores range from $\left \langle -1, 1 \right \rangle$, where -1 denotes a debunking and 1 a promoting stance towards the conspiracy.
    Only search results from queries that were executed both by the reference study and us are considered.
}
\small
\label{tab:hussein-comparison-search}
\begin{tabularx}{\textwidth}{lllp{22mm}X}
\toprule
Topic            & Hussein              & Ours   & Change & Inspection \\ \midrule

9/11             & -0.16                & -0.06 & No (n.s.d.)
& Smaller changes that depend on search query.  \\

Chemtrails       & -0.2                 & -0.47 & No (n.s.d.)
& Drop in promoting videos (from 45\% to 12\%) in 2 queries. \\

Flat earth       & -0.58                & -0.41 & No (n.s.d.)
& 2 queries worsen a lot due to new content. Other queries improve. \\

Moon landing     & -0.6                 & -0.59 & No (n.s.d.)
& Smaller decrease in number of neutral and increase of debunking videos. \\

Anti-vaccination & -0.8                 & -0.63 & \textcolor{red}{Worse} (U=324,~p=$1.3\mathrm{e}{-9}$)
& Drop in number of debunking and increase in number of neutral videos.  \\ \bottomrule
\end{tabularx}
\end{table*}

\begin{table*}
\caption{
    Comparison of normalized scores for up-next and top-5 recommendations with data from the reference study.
    Normalized scores range from $\left \langle -1, 1 \right \rangle$, where -1 denotes a debunking and 1 a promoting stance towards the conspiracy.
    Last 10 out of 20 watched videos in reference data are considered.
    Last 2 out of 40 watched videos in our data are considered.
}
\small
\label{tab:hussein-comparison-recommendations}
\begin{tabularx}{\textwidth}{llllX}
\toprule
Topic            & Hussein & Ours   & Change  & Inspection \\ \midrule
9/11             & 0.14    & 0.26  & No (n.s.d.)                   & Similar distribution, more promoting videos. \\
Chemtrails       & 0.05    & 0.03  & No (n.s.d.)                   & More neutral results. \\
Flat earth       & -0.16   & -0.15 & No (n.s.d.)                   & Similar distribution. \\
Moon landing     & -0.08   & -0.32 & \textcolor{DarkGreen}{Better} (U=2954.5, p=$8\mathrm{e}{-6}$)   & More debunking videos. \\
Anti-vaccination & -0.28   & 0     & \textcolor{red}{Worse} (U=664, p=$1.6\mathrm{e}{-9}$)       & Less debunking videos, more neutral and promoting. \\ \bottomrule
\end{tabularx}
\end{table*}

\subsection{RQ2: How does misinformation filter bubble effect change, when debunking videos are watched? }

\begin{table*}
\caption{
Comparison of SERP-MS scores for top-10 search results in promoting and debunking phase of our experiment.
Three points are compared: start of promoting phase (S1), end of promoting phase (E1), end of debunking phase (E2).
}
\small
\label{tab:comparison-search}
\begin{tabularx}{\textwidth}{p{12mm}p{11mm}p{44mm}X}
\toprule
Topic            & SERP-MS & Change & Inspection \\ \midrule
9/11             &
S1: -0.07 \newline E1: -0.06 \newline E2: -0.11
&
S1--E1: n.s.d. \newline
E1--E2: n.s.d. \newline
S1--E2: n.s.d.
&
E2: More debunking videos in one query (30\% instead of 12\% at S1 and 11\% at E1 in query ``9/11'').
\\ \hline
Chemtrails       &
S1: -0.45 \newline E1: -0.47 \newline E2: -0.49
&
S1--E1: n.s.d. \newline
E1--E2: n.s.d. \newline
S1--E2: n.s.d.
&
E2: The ``Chemtrail'' search query showed an increase in number of debunking videos (from 66\% at S1 and 69\% at E1 to 80\%) and a decrease in promoting (from 10\% to 0\%).
\\ \hline
Flat earth       &
S1: -0.27 \newline E1: -0.41 \newline E2: -0.45
&
S1--E1: \textcolor{DarkGreen}{better} (U=1737.5, p=0.0007) \newline
E1--E2: n.s.d. \newline
S1--E2: \textcolor{DarkGreen}{better} (U=1795.5, p=$1.6\mathrm{e}{-4}$)
&
E1: Change goes against expectations.
Promoting videos disappear in 3 search queries and decrease in another one (from 36\% to 30\%).
\newline
E2: Similar change as in E1 with a further decrease in promoting videos in one query (from 30\% to 22\%) and reordered videos in another.
\\ \hline
Moon landing     &
S1: -0.57 \newline E1: -0.57 \newline E2: -0.59
&
S1--E1: n.s.d. \newline
E1--E2: n.s.d. \newline
S1--E2: n.s.d.
&
E2: Reordered search results in ``moon hoax'' query---debunking videos moved higher.
\\ \hline
Anti-vacc. &
S1: -0.6 \newline E1: -0.63 \newline E2: -0.68
&
S1--E1: n.s.d. \newline
E1--E2: n.s.d. \newline
S1--E2: \textcolor{DarkGreen}{better} (U=1608.5, p=0.0003)
&
E2: Increase in debunking videos across multiple queries (from 60\% at S1 and 61\% at E1 to 67\%).
\\ \bottomrule
\end{tabularx}
\end{table*}

\begin{table*}
\caption{
Comparison of changes in average normalized scores for top-10 recommendations in promoting and debunking phase of our experiment.
Three points are compared: start of promoting phase (S1), end of promoting phase (E1), end of debunking phase (E2).
}
\small
\label{tab:comparison-recommendations}
\begin{tabularx}{\textwidth}{p{13mm}p{11mm}p{43mm}X}
\toprule
Topic            & Score & Change & Inspection \\ \midrule
9/11             &
S1: 0.1 \newline E1: 0.42 \newline E2: 0.07
&
S1--E1: \textcolor{red}{worse} (U=45.5, p=$2.6\mathrm{e}{-5}$) \newline
E1--E2: \textcolor{DarkGreen}{better} (U=372, p=$3\mathrm{e}{-6}$) \newline
S1--E2: n.s.d.
&
E1: Number of promoting videos increased (from 14\% to 43\%) and neutral videos decreased (from 83\% to 56\%).
\newline
E2: The numbers of promoting and neutral videos returned to levels comparable to start (13\% and 82\%).
\\ \hline
Chemtrails       &
S1: 0 \newline E1: 0.05 \newline E2: -0.15
&
S1--E1: n.s.d. \newline
E1--E2: \textcolor{DarkGreen}{better} (U=323, p=0.0006) \newline
S1--E2: \textcolor{DarkGreen}{better} (U=330, p=$1.6\mathrm{e}{-4}$)
&
E2: There is an increase in a number of debunking videos (from 0\% at S1 and 3\% at E1 to 19\%).
In return, we end up in a state that is better than at the start.
\\ \hline
Flat earth       &
S1: -0.17 \newline E1: -0.06 \newline E2: -0.47
&
S1--E1: n.s.d. \newline
E1--E2: \textcolor{DarkGreen}{better} (U=375, p=$2\mathrm{e}{-6}$) \newline
S1--E2: \textcolor{DarkGreen}{better} (U=347, p=$6.3\mathrm{e}{-5}$)
&
E2: Similar to the Chemtrails conspiracy, there is an increase in number of debunking videos (from 19\% at S1 and 16\% at E1 to 48\%).
\\ \hline
Moon landing     &
S1: -0.2 \newline E1: -0.4 \newline E2: -0.42
&
S1--E1: n.s.d. \newline
E1--E2: n.s.d. \newline
S1--E2: n.s.d.
&
E1: Mean normalized score changes against expectation and improves (but not significantly).
\\ \hline
Anti-vacc. &
S1: -0.1 \newline E1: 0.04 \newline E2: -0.37
&
S1--E1: \textcolor{red}{worse} (U=74.5, p=0.0016) \newline
E1--E2: \textcolor{DarkGreen}{better} (U=310, p=$2\mathrm{e}{-6}$) \newline
S1--E2: \textcolor{DarkGreen}{better} (U=307.5, p=$1.6\mathrm{e}{-4}$)
&
E1: Increase in number of promoting videos (from 2\% to 13\%).
\newline
E2: Increase in number of debunking videos (from 12\% at S1 and 9\% at E1 to 37\%) and disappearance of promoting (from 2\% at S1 and 13\% at E1 to 0\%).
\\ \bottomrule 
\end{tabularx}
\end{table*}

\begin{table*}
\caption{
Comparison of changes in average normalized scores for top-10 home page results in promoting and debunking phase of our experiment.
Three points are compared: start of promoting phase (S1), end of promoting phase (E1), end of debunking phase (E2).
}
\small
\label{tab:comparison-home-page}
\begin{tabularx}{\textwidth}{p{13mm}p{11mm}p{43mm}X}
\toprule
Topic            & Score & Change & Inspection \\ \midrule
9/11             &
S1: -0.15 \newline E1: 0.03 \newline E2: 0.0
&
S1--E1: \textcolor{red}{worse} (U=56.0, p=$6.6\mathrm{e}{-5}$) \newline
E1--E2: n.s.d. \newline
S1--E2: \textcolor{red}{worse} (U=73.5, p=$4.35\mathrm{e}{-4}$)
&
E1: Increase in number of promoting videos (from 3\% to 19\%), slight decrease in number of debunking (from 18\% to 16\%).
\newline
E2: Decrease in promoting (to 15\%) and slight decrease in debunking (to 15\%).
\\ \hline
Chemtrails       &
S1: -0.23 \newline E1: -0.06 \newline E2: -0.4
&
S1--E1: \textcolor{red}{worse} (U=46.5, p=$2.2\mathrm{e}{-5}$) \newline
E1--E2: \textcolor{DarkGreen}{better} (U=396.0, p=0.0) \newline
S1--E2: \textcolor{DarkGreen}{better} (U=351.0, p=$2.9\mathrm{e}{-5}$)
&
E1: Increase in number of promoting videos (from 2\% to 10\%), and decrease in number of debunking (from 26\% to 16\%).
\newline
E2: Decrease in promoting (to 1\%) and increase in debunking (to 41\%).
\\ \hline
Flat earth       &
S1: -0.11 \newline E1: -0.06 \newline E2: -0.38
&
S1--E1: n.s.d. \newline
E1--E2: \textcolor{DarkGreen}{better} (U=372.5, p=$2\mathrm{e}{-6}$) \newline
S1--E2: \textcolor{DarkGreen}{better} (U=361.5, p=$1\mathrm{e}{-5}$)
&
E1: Increase in number of promoting videos (from 2\% to 10\%), and also in number of debunking (from 12\% to 16\%).
\newline
E2: Decrease in promoting (to 4\%) and increase in debunking (to 41\%).
\\ \hline
Moon landing     &
S1: -0.15 \newline E1: -0.15 \newline E2: -0.34
&
S1--E1: n.s.d. \newline
E1--E2: \textcolor{DarkGreen}{better} (U=354.5, p=$1.7\mathrm{e}{-5}$) \newline
S1--E2: \textcolor{DarkGreen}{better} (U=318.5, p=$1.2\mathrm{e}{-3}$)
&
E1: No change in number of promoting videos (at 2\%), and an increase in number of debunking (from 16\% to 18\%).
\newline
E2: Same number of promoting (2\%) and a further increase in debunking (to 36\%).
\\ \hline
Anti-vacc. &
S1: -0.22 \newline E1: -0.26 \newline E2: -0.53
&
S1--E1: n.s.d. \newline
E1--E2: \textcolor{DarkGreen}{better} (U=137.5, p=$3.6\mathrm{e}{-3}$) \newline
S1--E2: \textcolor{DarkGreen}{better} (U=169.0, p=$1.7\mathrm{e}{-4}$)
&
E1: Increase in number of promoting videos (from 4\% to 10\%), and also in number of debunking (from 25\% to 36\%).
\newline
E2: Decrease in promoting (to 3\%) and increase in debunking (to 57\%).
\\ \bottomrule 
\end{tabularx}
\end{table*}

Answering this research question requires four comparisons, each related to one of our hypotheses:
\begin{enumerate}
    \item comparison of metrics between start of promoting phase (S1) and end of promoting phase (E1) answering H2.0,
    \item comparison of metrics between end of promoting phase (E1) and end of debunking phase (E2) answering H2.1,
    \item comparison of metrics between start of promoting phase (S1) and end of debunking phase (E2) answering H2.2,
    \item comparison of the slope of metrics in the promoting phase and in the debunking phase answering H2.3.
\end{enumerate}

As already noted, automatically generated annotations using the trained ML model were used in addition to the manually labeled data for evaluating the comparisons on home page results and in case of comparison (4), also on top-10 recommendations.

\subsubsection{Comparison (1)}

There are changes in search results, recommendations and home page results after watching promoting videos (E1) compared to the start of the experiment (S1).
If there was a misinformation bubble effect developed, we would expect the metrics to worsen due to watching promoting videos (H2.0).
Regarding search results, the distribution of SERP-MS scores between S1 and E1 is indeed significantly different (MW U=34118.5, p-value=0.028).
However, the score actually improves---mean SERP-MS score changed from -0.39 (std 0.28) to -0.42 (std 0.3).
Table~\ref{tab:comparison-search} shows the change for individual topics.
Only the flat earth conspiracy shows significant differences and improved the SERP-MS score due to a decrease in promoting and an increase of debunking videos.

Top-10 recommendations also change their distribution of normalized scores significantly at E1 compared to S1 (MW U=4085, p-value=0.0397).
We observe that the mean normalized score worsens from -0.07 (std 0.24) to 0.01 (std 0.31).
Looking at individual topics in Table~\ref{tab:comparison-recommendations}, we can see that the change is significant in topics 9/11 and anti-vaccination that gain more promoting videos.

The overall change in home page results across all topics is statistically significantly different as well (MW U=3412.5, p-value=0.0002).
It worsens from -0.17 (std 0.14) to -0.09 (std 0.17).
When looking at the individual topics, we see statistically significant changes on home page in certain topics---9/11, and chemtrails both get worse (cf. Table~\ref{tab:comparison-home-page}). We see an increase in the proportion of promoting videos also in the flat earth and anti-vaccination topics, but these changes are not statistically significant. Interestingly, home page results in the moon landing topic see a higher proportion of debunking videos (although this change is also not significant).

\subsubsection{Comparison (2)} 
When examining the changes in search results, recommendations and home page results between the end of promoting phase (E1) and the end of debunking phase (E2), we expect the metrics would improve due to watching debunking videos, i.e., that we would observe misinformation bubble bursting (lessening of a misinformation filter bubble effect; H2.1).
However, SERP-MS scores in search results between E1 and E2 are not from statistically significantly different distributions, which is consistent with the fact that we did not observe misinformation bubble creation in search results in the first place.
Table~\ref{tab:comparison-search} shows minor improvements in SERP-MS scores across the topics, but they are not statistically significant.

Top-10 recommendations show more considerable differences and their overall distribution is significantly different when comparing E1 and E2 (MW U=7179.5, p-value=$1.8\mathrm{e}{-9}$).
Mean normalized score improves from 0.01 (std 0.31) to -0.27 (std 0.27).
Table~\ref{tab:comparison-recommendations} shows significantly different distributions for all topics except for moon landing conspiracy.
All topics show an improvement in normalized scores.
The 9/11 topic shows a decrease in promoting videos, while the other topics show an increase in the number of debunking videos.

Home page results also show an overall significantly different distribution of the normalized scores between E1 and E2 (MW U=6876.5, p-value=$0.0$).
There are statistically significant improvements in all topics except for 9/11.
All topics except for moon landing show a decrease in the number of promoting videos and all topics except for 9/11 show a rise in debunking videos.

\subsubsection{Comparison (3)} 
When examining the differences between the start (S1) and end of the experiment (E2), we expect the metrics would improve due to watching debunking videos despite watching promoting videos before that (H2.2).
The distribution of SERP-MS scores in search results is statistically significantly different when comparing S1 and E2 (MW U=36515, p-value=0.0002).
Overall, we see an improvement in mean SERP-MS score from -0.39 (std 0.28) to -0.46 (std 0.29).
Table~\ref{tab:comparison-search} shows that only topics flat earth and anti-vaccination significantly changed their distributions, but all topics show an improvement according to our expectations.
The improvement is due to increases in debunking videos, decreases in promoting videos, or reordered search results in some search queries.

Similarly, top-10 recommendations at E2 come from a significantly different distribution than at S1 (MW U=6940.5, p-value=$2.9\mathrm{e}{-7}$).
Mean normalized score improves from -0.07 (std 0.24) to -0.27 (std 0.27).
Table~\ref{tab:comparison-recommendations} shows a significant difference in distributions for all topics except for 9/11 and moon landing conspiracies.
Mean normalized scores improve compared to S1 in all topics except for 9/11, where the score is comparable to S1 (due to the numbers of promoting and neutral videos returning to S1 levels).
Other topics show increases in the numbers of debunking videos.

Home page results at E2 also come from a statistically significantly different distribution compared to S1 (MW U=6340.0, p-value=$0.0$).
All topics except for 9/11, which interestingly gets worse, show a statistically significant improvement in the metrics most commonly due to an increase in the number of debunking videos.

\begin{table*}
\caption{
Difference to expected linear trend (\textbf{DIFF-TO-LINEAR} metric) across top-10 recommendations (``Recomm.''), and home page results (``Home'') in the promoting phase (phase 1), and debunking phase (phase 2) for topics with statistically significant changes in the normalized score metrics in the respective phases (cells with ``-'' denote cases without statistically significant changes). Positive values in the promoting phase indicate that normalized score worsens faster than linearly and negative values in the debunking phase indicate that it improves faster than linearly. The promoting phase shows smaller differences to the expected linear trend compared to the debunking phase. On the other hand, normalized score improves much faster than linear trend in the debunking phase in most cases.
}
\small
\label{tab:diff-to-linear}
\begin{tabularx}{\textwidth}{llcccccX}
\toprule
Phase  & Modality & 9/11   & Chemtrails & Flat earth & Moon land. & Anti-vacc. & Inspection \\ \midrule
1 & Recomm. & \textcolor{red}{0.55} & - & - & - & \textcolor{DarkGreen}{-1.631} & Worsens faster than linearly in case of 9/11 and slower than linearly in case of anti-vaccination. \\
  & Home & \textcolor{DarkGreen}{-0.268} & \textcolor{DarkGreen}{-0.016} & - & - & - & Close to linear changes. \\ \hline
2 & Recomm. & \textcolor{DarkGreen}{-3.255} & \textcolor{DarkGreen}{-8.645} & \textcolor{DarkGreen}{-6.555} & - & \textcolor{DarkGreen}{-11.454} & Fast improvement. \\
  & Home & - & \textcolor{DarkGreen}{-7.735} & \textcolor{DarkGreen}{-11.935} & \textcolor{DarkGreen}{-7.092} & \textcolor{DarkGreen}{-3.923} & Fast improvement. \\ \bottomrule
\end{tabularx}
\end{table*}

\begin{figure*}
\centering
\includegraphics[width=\textwidth]{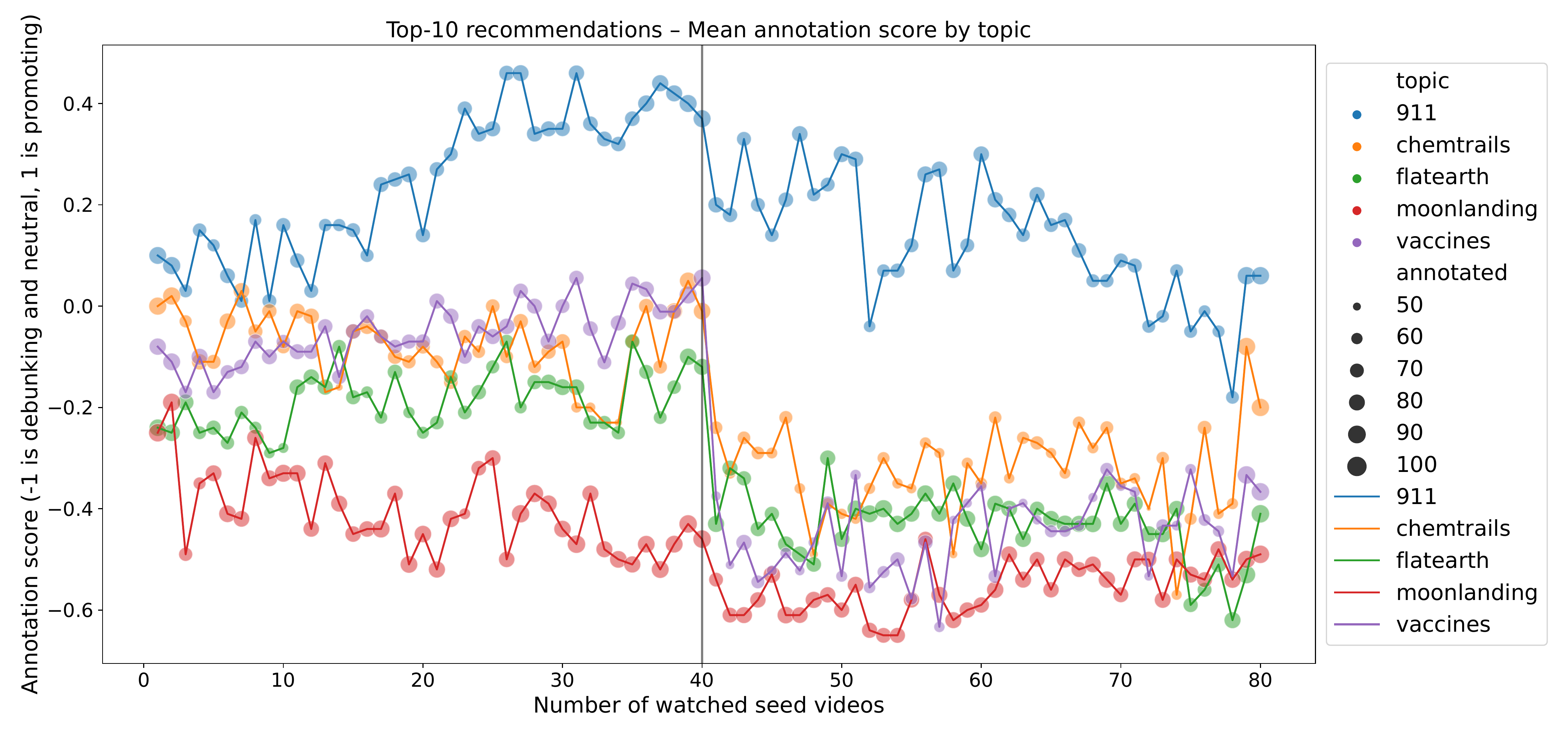}
\includegraphics[width=\textwidth]{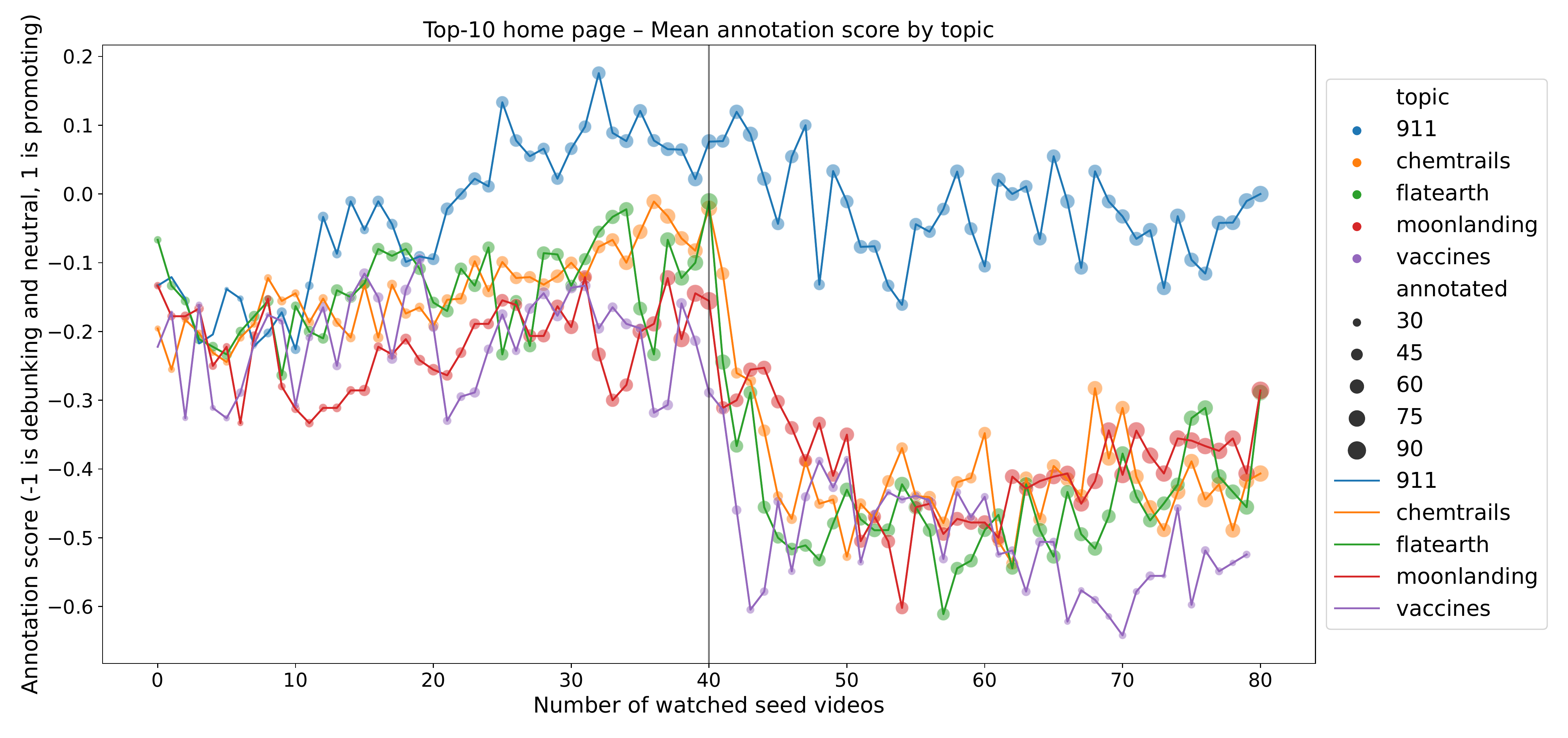}
\caption{Changes in average normalized score for top-10 recommendations (top) and for home page results (bottom) over the duration of the experiment. The normalized score ranges from -1 for all debunking to +1 for all promoting recommendations. The X-axis shows the number of videos that the agents had watched before the recorded recommendations. Recall that the agents first watched 40 promoting and then 40 debunking videos. For some topics, one can observe a sudden drop in the normalized score after the 40th video, i.e., when agents started watching debunking videos. As some of the video labels are generated by the trained machine learning model, we also show the proportion of manually annotated videos out of all recommendations using the size of dots.}
\label{fig:annotation-score-by-topic}
\end{figure*}

\begin{figure*}
\centering
\includegraphics[width=\textwidth]{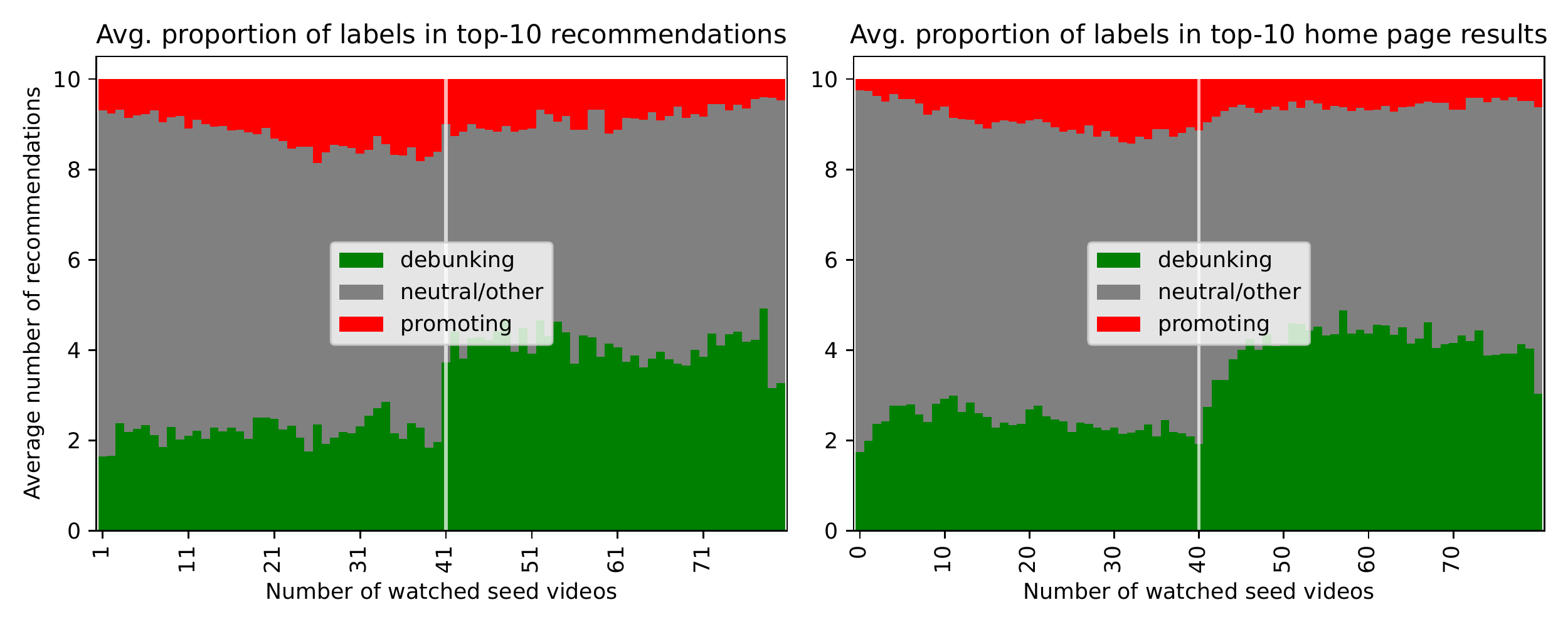}
\caption{Proportions of stance labels (promoting, debunking, neutral) of videos in top-10 recommendations (left) and home page results (right) over the duration of the experiment.}
\label{fig:proportions-of-recommendations}
\end{figure*}

\subsubsection{Comparison (4)} 

Finally, we want to look deeper at the change in the metrics throughout the experiment. Our interest is in evaluating the slope of the misinformation normalized score and we expect it to increase approximately linearly as the 40 promoting videos are watched and decrease approximately linearly as the 40 debunking videos are watched (H2.3). We use the \textbf{DIFF-TO-LINEAR} metric defined in Section~\ref{sec:methodology:metrics} and evaluate it for top-10 recommendations and home page results within topics that showed statistically significant changes in the normalized scores (in S1 vs. E1 and in E1 vs. E2 comparisons). Table~\ref{tab:diff-to-linear} shows the results. In most cases, we can see that the change is faster than linear---in the promoting phase, recommendations in the 9/11 topic show positive values. This indicates that they worsen faster than linearly. The remaining topics show negative values that are close to 0 (except for anti-vaccination which worsens slower than linearly). The change is larger in the debunking phase---all topics show faster improvement (negative values) of top-10 recommendations and home page results.
Figure~\ref{fig:annotation-score-by-topic} lets us look at these changes in normalized score in more detail. We can observe the change that happens right after the end of promoting phase---there is a sudden decrease (improvement) in the score. This is visible for both top-10 recommendations and home page results in most topics. The main exception is the 9/11 topic that shows more gradual changes compared to other topics both in the promoting and debunking phase.
To look even deeper at how the proportions of promoting, debunking, and neutral videos change over the experiment, we can refer to Figure~\ref{fig:proportions-of-recommendations}. Here we can see a sudden increase in the number of debunking videos especially in recommendations at the start of the debunking phase. Proportion of promoting videos increases gradually over the promoting phase and decreases over the debunking phase.
\section{Discussion and Conclusions}
\label{sec:conclusions}

In the paper, we presented an audit of misinformation present in search results, recommendations, and home page results on the video-sharing platform YouTube. To support reproducibility\footnote{Note that the reproducibility may suffer to some extent due to the dynamic nature of the platform, where some of the videos we used for seeding or encountered may no longer be available, as we discuss in more detail below.}, we publish the collected data and source codes for the experiment.

We aimed at verifying a hypothesis that there is less misinformation present in both search results and recommendations as a result of the ongoing improvements in YouTube recommender systems and policies~\cite{YouTube2020policies,YouTube2021perspective} (H1.1). The comparison was done against a study carried out in mid-2019 by Hussein et al.~\cite{Hussein2020}. We were interested, whether we could still observe the misinformation bubble effect after watching videos promoting conspiracy theories (H2.0). In addition to the previous studies, we also examined bubble bursting behavior. Namely, we aimed to verify whether misinformation bubbles could be burst if we watched videos debunking conspiracy theories (H2.1). We also hypothesized that watching debunking videos (even after a previous sequence of promoting videos) would still decrease the amount of misinformation compared to the initial state with no watch history at the start of the study (H2.2).
Finally, we investigated the dynamics of change in the prevalence of misinformation promoting videos and hypothesized that their number would increase linearly (using linear change as a baseline) as misinformation promoting videos are watched, and decrease linearly as more and more misinformation debunking videos are watched (H2.3).

Regarding hypothesis H1.1, we did not find a significantly different amount of misinformation in search results in comparison to the reference study. A single topic (anti-vaccination) showed a statistically significant difference. However, it did not agree with the hypothesis as the metric \emph{worsened} due to more neutral and less debunking videos. Recommendations showed significant differences across multiple topics but were not significantly different overall. A single topic (moon landing) improved normalized scores of recommendations in agreement with the hypothesis. Yet, the anti-vaccination topic worsened its scores.
We suspect the changes in search results and recommendations were influenced mostly by changes in content, but currently it is not possible to reliably differentiate between endogenous (e.g., change in the recommendation algorithm or platform policy) and exogenous changes (e.g., change in content on the platform). Overall, our results did not show a significant improvement in the prevalence of misinformation in the audited topics, although it is worth noting that the absolute numbers of misinformation promoting videos are relatively low in most of the audited topics.

We did not observe the development of misinformation filter bubble effect in search results (H2.0) despite watching promoting videos. On the other hand, recommendations behaved according to our hypothesis, and their overall normalized scores worsened. By making use of predictions from a trained ML model, we evaluated home page results as well. They showed a statistically significant change in their overall distribution; the scores worsened in 9/11 and chemtrails topics. Since we did not observe a filter bubble effect in search results, we did not observe any bubble bursting effect there either. Results did not show a statistically significant difference between the end of promoting phase and the end of the debunking phase. Recommendations as well as home page results showed more considerable differences that were statistically significant and confirmed the hypothesis H2.1. We showed that watching debunking videos decreases the number of misinformation promoting videos and increases the number of misinformation debunking videos in search results, recommendations and home page results when compared to the initial state (start of the experiment), which confirms our hypothesis H2.2. We observed an improvement of SERP-MS scores for search results and normalized scores for recommendations and home page results in most topics.
Finally, we inspected the trend of change in metrics over the course of the experiment as we expected them to change linearly as misinformation promoting or debunking videos are watched (H2.3). We saw large deviations from a linear trend in the debunking phase. There was a sudden improvement of the misinformation score after the first watched debunking video caused by an increase in the number of debunking videos. Although changes in the score continue as more videos are watched, there is a strong contextuality of recommendations and home page results with the most recently watched videos.

Based on our results, we can conclude that users, even with a watch history of promoting conspiracy theories, do not get enclosed in a misinformation filter bubble \emph{when they search} on YouTube. However, we do observe this effect in video recommendations and home page results with varying degrees depending on the topic. At the same time, \emph{watching debunking videos helps practically in all cases} to decrease the amount of misinformation that the users see. Additionally, although we expected to see less misinformation than the previous studies reported, this was in general not the case. Worsening in the anti-vaccination topic was partially expected due to the COVID-19 pandemic. However, it is interesting that we also observed a worse situation in the 9/11 topic. In fact, this topic served as a sort of a gateway to misinformation videos on other topics.

Besides results, several limitations of our study need to be pointed out. First, we investigated only a limited amount of topics---these did not include, for example, recent QAnon conspiracy, disinformation and propaganda related to Russia-Ukraine war, or COVID-19 related conspiracies (which were present only indirectly through anti-vaccination narratives). However, our topics were explicitly selected to allow comparison with the reference study. Additional audit studies like ours are needed to investigate misinformation prevalence in constantly emerging new topics. It would also be interesting to extend the audit for different geolocations (other than USA) and languages other than English and to select topics relevant for these new contexts. However, the substantial required effort (especially with respect to the data collection and following manual annotation) is currently a serious limiting factor in this regard. Another limitation is that we included only a limited set of agent interactions with the platform (search and video watching). Real users also like or dislike videos, subscribe to channels, leave comments or click on the search results or recommendations. A more human-like bot simulation, with these interactions and possible inclusion of human biases bursting remains a subject of the future work.  Furthermore, as already stated, it is currently difficult (if even possible at all) to distinguish between endogenous and exogenous factors that impact the observed results. The direct comparison between independent audit studies is also sometimes difficult due to many factors that might confound the results and that the researchers must be aware of when interpreting the observed differences.

Despite the limitations, our audit showed that YouTube (similar to other platforms), despite their best efforts so far, can still promote misinformation seeking behavior to some extent. Even a year after running the experiment, the majority of videos that we annotated as promoting (either those we used as seed videos, or the additional ones we encountered) still remain on the platform. Only 64 out of the 449 promoting videos (\(\sim14\%\)) are no longer available at the platform. However, we can see the focus on dealing with misinformation from the platform, as the number of removed promoting videos is higher than removed debunking (8 out of 760 seed and encountered) and neutral (41 out of 2,042 seed and encountered) videos, so the removal cannot be attributed solely to the evolution of videos on the platform (e.g., how many are created and removed regardless of topic). We can also see the focus on major misinformation topics, as 75\% of the removed promoting seed videos come from the anti-vaccination topic and from the overall promoting videos (seed and encountered) that were removed, \(\sim49\%\) come from the anti-vaccination and \(\sim15\%\) come from the flat earth topic. The relatively low number of removed promoting videos can be a result of a difference between what we consider to be a promoting video and what is considered to be a misinformation video by the platform, although the number of potentially harmful videos present on the platform is still high. 

The results also motivate the need for independent continuous and automatic audits of YouTube and other social media platforms~\cite{Simko2021}, since we observed that the amount of misinformation in a topic could change over time due to endogenous as well as exogenous factors. The partial use of automated annotation of recommended videos shown in this paper is a step towards this goal. However, it is crucial for any automated approach to be trustworthy, i.e., to be robust (e.g., against potential concept drifts), explainable, and to include humans in the loop (or better yet, in control) at all stages of the audit. This remains our future work.

\begin{acks}
This work was partially supported by The Ministry of Education, Science, Research and Sport of the Slovak Republic under the Contract No. 0827/2021; by the Central European Digital Media Observatory (CEDMO), a project funded by the European Union under the Contract No. 2020-EU-IA-0267; by TAILOR, a project funded by EU Horizon 2020 research and innovation programme under GA No. \href{https://doi.org/10.3030/952215}{952215}; and by vera.ai, a project funded by the European Union under the Horizon Europe, GA No. \href{https://doi.org/10.3030/101070093}{101070093}.
\end{acks}

\bibliographystyle{ACM-Reference-Format}
\bibliography{references}

\end{document}